\documentclass[11pt,aps,prd,preprint,groupedaddress,tightenlines,nofootinbib,floatfix]{revtex4}
\usepackage[dvipdfm]{graphicx}

\usepackage{amssymb}
\usepackage{amsmath}
\usepackage{epstopdf}
\usepackage{booktabs}

\makeatletter
\newcommand{\figcaption}[1]{\def\@captype{figure}\caption{#1}}
\newcommand{\tblcaption}[1]{\def\@captype{table}\caption{#1}}
\makeatother

\def\simge{\mathrel{%
       \rlap{\raise 0.511ex \hbox{$>$}}{\lower 0.511ex \hbox{$\sim$}}}}
\def\simle{\mathrel{
       \rlap{\raise 0.511ex \hbox{$<$}}{\lower 0.511ex \hbox{$\sim$}}}}

\begin{document}

\title{Latent heat at the first order phase transition point of SU(3) gauge theory}

\author{Mizuki Shirogane$^{1}$, Shinji Ejiri$^{2}$, Ryo Iwami$^{1}$, Kazuyuki Kanaya$^{3,4}$, 
Masakiyo Kitazawa$^5$ \\
(WHOT-QCD Collaboration)
}
\affiliation{
$^1$Graduate School of Science and Technology, Niigata University, Niigata 950-2181, Japan\\
$^2$Department of Physics, Niigata University, Niigata 950-2181, Japan\\
$^3$Center for Integrated Research in Fundamental Science and Technology (CiRfSE), University of Tsukuba, Tsukuba, Ibaraki 305-8571, Japan\\
$^4$Faculty of Pure and Applied Sciences, University of Tsukuba, Tsukuba, Ibaraki 305-8571, Japan\\
$^5$Department of Physics, Osaka University, Toyonaka, Osaka 560-0043, Japan
}

\date{May 9, 2016}

\begin{abstract}
We calculate the energy gap (latent heat) and pressure gap between the hot and cold phases of the SU(3) gauge theory at the first order deconfining phase transition point.
We perform simulations around the phase transition point with the lattice size in the temporal direction $N_t=6,$ $8$ and $12$ and extrapolate the results to the continuum limit. 
We also investigate the spatial volume dependence.
The energy density and pressure are evaluated by the derivative method with non-perturabative anisotropy coefficients. 
We adopt a multi-point reweighting method to determine the anisotropy coefficients.  
We confirm that the anisotropy coefficients approach the perturbative values as $N_t$ increases. 
We find that the pressure gap vanishes at all values of $N_t$ when the non-perturbative anisotropy coefficients are used.
The spatial volume dependence in the latent heat is found to be small on large lattices. 
Performing extrapolation to the continuum limit, we obtain 
$
\Delta \epsilon/T^4 = 0.75 \pm 0.17
$
and
$
\Delta (\epsilon -3 p)/T^4 = 0.623 \pm 0.056.
$
\end{abstract}

\maketitle

\section{Introduction}
\label{sec:intro}

Determination of the equation of state from a first principle calculation of QCD is one of the most important topics in the study of the quark matter \cite{YHM}.
In this paper, we study thermodynamic quantities around the first order deconfining phase transition in the SU(3) gauge theory (the quenched approximation of QCD).
First order phase transitions are expected in the high density region of QCD and also in the many-flavor QCD aiming at construction of a walking technicolor model \cite{Appelquist:1995en,Kikukawa:2007zk,Ejiri:2012rr}.
The SU(3) gauge theory at finite temperature provides us with a good testing ground to study characteristic features of first order transition and to develop techniques to investigate thermodynamic quantities around it.

At a first order phase transition point, two phases coexist at the same time.
To keep the balance between them, the pressure must be the same in the two phases.
On the other hand, the energy density is different in these phases.
The difference is the latent heat which is one of the most important physical quantities characterizing the first order phase transition.
In numerical studies of QCD, the integral method is widely adopted \cite{integral}. 
However, in the integral method, the pressure gap is set to be zero in the formulation.
To study the pressure gap itself, we adopt the derivative method in this study. 

In the derivative method, the values of the derivatives of gauge coupling 
constants with respect to the anisotropic lattice spacings, 
which we call the anisotropy coefficients, are required.
The anisotropy coefficients in SU(3) gauge theory have been calculated in the lowest order 
perturbation theory by Karsch \cite{karsch}.
However, the perturbative coefficients are known to lead to
pathological results such as negative pressure 
and non-vanishing pressure gap at the deconfining transition point,
when the lattice size in the temporal direction $N_t$ is small. 
This motivated a non-perturbative calculation of the anisotropy coefficients of Ref.~\cite{ejiri98}, 
in which the pressure gap using the  non-perturbative anisotropy coefficients is confirmed to 
vanish at the first order phase transition on lattices with $N_t=4$ and $6$.
The latent heat was also computed using the non-perturbative anisotropy coefficients at  $N_t=4$ and 6. 
We now extend the study to larger values of $N_t$ to carry out the continuum extrapolation. 
We also adopt larger spatial volumes and study the spatial volume dependence of the results.

In the next section, we introduce the basic formulation and the methods to study the energy density and pressure by the derivative method.
The non-perturbative anisotropy coefficients are calculated by the method proposed in Ref.~\cite{ejiri98}.  
A multi-point reweighting method is used for the calculation of the expectation values of the plaquette as well as the Polyakov loop and its susceptibility at the phase transition point.
The results of our numerical simulation is given in Sec.~\ref{sec:results}:
Our simulation parameters are summarized in Sec.~\ref{sec:parameter}.
The results of the anisotropy coefficients are shown in Sec.~\ref{sec:aniso}.
The separation of the configurations into the hot and cold phases are discussed in Sec.~\ref{sec:phase}.
We then compute the latent heat and the pressure gap, and evaluate the latent heat in the continuum limit in Sec.~\ref{sec:latent}.
Our conclusion and outlook are summarized in Sec.~\ref{sec:conclusion}.

\section{Method}
\label{sec:method}

\subsection{Latent heat and pressure gap}
\label{sec:eos}

The energy density $\epsilon$ and the pressure $p$
are defined by the derivatives of the partition function $Z$ 
in terms of the temperature $T$ 
and the physical volume $V$ of the system
\begin{eqnarray}
\epsilon = - \frac{1}{V} \left. \frac{\partial \ln Z}{\partial \,T^{-1}} \right|_{V}, 
\hspace{5mm} 
 p       = T \left. \frac{\partial \ln Z}{\partial \, V} \right|_{T}. 
\label{eqn:ep}
\end{eqnarray}
On a lattice with a size $N_s^3\times N_t$, the volume and temperature are given by 
$V = (N_s a_s)^3$ and $T = 1 / (N_t a_t)$, 
with $a_s$ and $a_t$ the lattice spacings in spatial and temporal directions.
Because $N_s$ and $N_t$ are discrete parameters, 
the partial differentiations in Eq.~(\ref{eqn:ep}) are performed 
by varying $a_s$ and $a_t$ independently on anisotropic lattices \cite{satz,karsch}.
The anisotropy on a lattice is realized by introducing 
different coupling parameters in temporal and spatial directions.
For an SU($N_c$) gauge theory, the standard plaquette action 
on an anisotropic lattice is given by
\begin{eqnarray}
 S = -\beta_s \sum_{i<j \ne 4} \sum_{x} P_{ij}(x) 
     -\beta_t \sum_{i \ne 4} \sum_{x} P_{i4}(x),
\end{eqnarray}
where
$ P_{\mu \nu}(x) = N_{c}^{-1} {\rm Re \ Tr} 
[ U_{\mu}(x) U_{\nu}(x+\hat{\mu})
   U^{\dagger}_{\mu}(x+\hat{\nu}) U^{\dagger}_{\nu}(x) ]$
is the plaquette in the $(\mu,\nu)$ plane.
With this action, 
the energy density and pressure are given by 
\begin{eqnarray}
\epsilon &=& - \frac{3 N_t^{4} T^{4}}{\xi^3} 
  \left\{ \left(a_t \frac{\partial \beta_s}{\partial a_t} 
 -\xi \frac{\partial \beta_s}{\partial \xi}\right) 
 \left(\langle P_s \rangle - \langle P \rangle_0\right) + 
  \left(a_t \frac{\partial \beta_t}{\partial a_t} 
 -\xi \frac{\partial \beta_t}{\partial \xi}\right) 
 \left(\langle P_t \rangle - \langle P \rangle_0\right) \right\}, \label{enrg} \\
p &=& \frac{N_t^{4} T^{4}}{\xi^3} 
  \left\{\xi \frac{\partial \beta_s}{\partial \xi} \,
 \left(\langle P_s \rangle  - \langle P \rangle_0\right)
 +\xi \frac{\partial \beta_t}{\partial \xi} \,
 \left(\langle P_t \rangle - \langle P \rangle_0\right) \right\}, \label{prs}
\end{eqnarray}
where $\langle P_{s(t)} \rangle$ is
the space(time)-like plaquette expectation value, 
\begin{eqnarray}
P_s = \frac{1}{3N_{\rm site}} \sum_{i<j \ne 4} \sum_x P_{ij}(x) 
\hspace{5mm} {\rm and} \hspace{5mm} 
P_t = \frac{1}{3N_{\rm site}} \sum_{i \ne 4} \sum_x P_{i4}(x),
\end{eqnarray}
and $\langle P \rangle_0$ is the plaquette expectation value 
on a zero temperature lattice.
These expectation values can be computed by numerical simulations of the SU($N_c$) gauge theory non-perturbatively.
Here, for later convenience, we have chosen $a_t$ and 
$\xi \equiv a_s/a_t$ as 
independent variables to vary the lattice spacings, 
instead of $a_s$ and $\xi$ as adopted in Ref.~\cite{karsch}.

The derivatives of the gauge coupling constants
with respect to the anisotropic lattice spacings 
\begin{eqnarray}
a_t \frac{\partial \beta_s}{\partial a_t},
\hspace{5mm}
a_t \frac{\partial \beta_t}{\partial a_t},
\hspace{5mm}
\frac{\partial \beta_s}{\partial \xi},
\hspace{5mm}
\frac{\partial \beta_t}{\partial \xi},
\end{eqnarray}
are called the anisotropy coefficients. 
They are computed from a requirement that the effects of anisotropy in the physical 
observables can be absorbed by a renormalization of the coupling parameters.
The anisotropy coefficients do not depend on the temperature, 
because the renormalization is independent of the temperature.
To calculate the energy density and pressure by a simulation on isotropic lattices, 
we need the values of anisotropy coefficients at $\xi=1$. 

Performing simulation  at the transition temperature with $\xi=1$,
the differences of the energy density and pressure between hot and cold phases, 
i.e. the latent heat $\Delta \epsilon$ and pressure gap $\Delta p$, 
can be calculated by separating the configurations into the hot and cold phases,
\begin{eqnarray}
\frac{\Delta \epsilon}{T^4} &=& - 3 N_t^{4} 
  \left\{ \left(a_t \frac{\partial \beta_s}{\partial a_t} 
 - \frac{\partial \beta_s}{\partial \xi}\right) 
 \left( \langle P_s \rangle_{\rm hot} - \langle P_s \rangle_{\rm cold} \right) + 
  \left(a_t \frac{\partial \beta_t}{\partial a_t} 
 - \frac{\partial \beta_t}{\partial \xi}\right) 
 \left( \langle P_t \rangle_{\rm hot} - \langle P_t \rangle_{\rm cold} \right) \right\}, \ \ 
\label{eq:denrg} \\
\frac{\Delta p}{T^4} &=& N_t^{4}  
  \left\{ \frac{\partial \beta_s}{\partial \xi} \,
 \left( \langle P_s \rangle_{\rm hot}  - \langle P_s \rangle_{\rm cold} \right)
 + \frac{\partial \beta_t}{\partial \xi} \,
 \left( \langle P_t \rangle_{\rm hot} - \langle P_t \rangle_{\rm cold} \right) \right\}, 
\label{eq:dprs}
\end{eqnarray}
where $\langle \cdots \rangle_{\rm hot}$ and $\langle \cdots \rangle_{\rm cold}$ mean the expectation values in the hot and cold phases, respectively.
Separation of the configurations into the phases will be discussed in Sec.~\ref{sec:phase}.
Note that, in the calculations of $\Delta \epsilon$ and $\Delta p$, the zero temperature subtraction is not necessary. 
In the next subsection, we discuss that the anisotropy coefficients can be calculated 
by the same finite temperature simulations around the transition point on isotropic lattices \cite{ejiri98}.

\subsection{Anisotropy coefficients}
\label{sec:anisotropy}

We compute the anisotropy coefficients non-perturbatively 
following Ref.~\cite{ejiri98}.
This method is based on the measurement of the phase transition line in the $(\beta_s, \beta_t)$ plane.
On the transition line, the temperature $T=(N_t a_t)^{-1}$ is constant, thus $a_t$ is constant. 
From this information, one can determine the anisotropy coefficients.

Another non-perturbative way to determine the anisotropy coefficients is the so-called  ``matching method''
\cite{burgers,fujisaki,scheideler,engels00,klassen}.
In this method, one first determines $\xi$ as a function of $\beta_s$ 
and $\beta_t$ by matching space-like and time-like Wilson 
loops on anisotropic lattices, 
and then numerically determines $\partial\gamma/\partial\xi$
at $\xi=1$, where $\gamma = \sqrt{\beta_t/\beta_s}$.
Interpolation of the Wilson loop data at different sizes 
or interpolation of $\xi$ at different $\gamma$ using 
an appropriate ansatz is required to evaluate $\partial\gamma/\partial\xi$.
The method of Ref.~\cite{ejiri98} avoids uncertainties due to such interpolations. 

On isotropic lattices with $a_s=a_t=a$ and $\xi =1$, the coupling constants satisfy $\beta_s = \beta_t \equiv \beta$
and we have 
\begin{eqnarray}
\left( a_t \frac{\partial \beta_s}{\partial a_t} \right)_{\xi = 1}
= \left( a_t \frac{\partial \beta_t}{\partial a_t} \right)_{\xi = 1}
= a \frac{d \beta}{d a} 
= 2N_{c} \, a \frac{d g^{-2}}{d a},
\end{eqnarray}
where 
$\beta = 2N_c\,g^{-2}$ and 
$\displaystyle{a \frac{d g^{-2}}{d a}}$ is 
the beta function at $\xi = 1$,
whose non-perturbative value is well studied by numerical simulations of the SU(3) gauge theory \cite{taro,boyd,edwards}. 
See also Refs.~\cite{Guagnelli:1998ud,Necco:2001xg,Asakawa:2015vta,francis15} for determination of the lattice scale.  
Moreover, a combination of the remaining two anisotropy coefficients 
is known to be related to the beta function \cite{karsch} as%
\footnote{
In \cite{karsch}, a corresponding equation is given for
$(\partial \beta_{s (t)} / \partial \xi )_{a_s: {\rm fixed}}$.
}
\begin{eqnarray}
\left(\frac{\partial \beta_s}{\partial \xi} 
+ \frac{\partial \beta_t}{\partial \xi}\right)
_{a_t : {\rm fixed},\, \xi = 1}  
= \frac{3}{2} \, a \frac{d \beta}{d a}. 
\label{eq:cubicsym}
\end{eqnarray}
This equation is derived by the following way.
The string tension $\sigma$ defined as
\begin{eqnarray}
\sigma a_s a_t = - \lim_{A \to \infty} \frac{1}{A} \ln \langle W_t \rangle,
\hspace{5mm}
\sigma a_s^2 = - \lim_{A \to \infty} \frac{1}{A} \ln \langle W_s \rangle,
\end{eqnarray}
is independent of $\xi=a_s/a_t$.
Here, $\langle W_s \rangle$ and $\langle W_t \rangle$ are the expectation values of space-like and time-like planer Wilson loop operators, respectively.
$A$ is the number of plaquettes enclosed by the Wilson loop. 
We then obtain 
\begin{eqnarray}
\frac{\partial (\sigma a_t^2)}{\partial \xi} &=& - \lim_{A \to \infty} \frac{1}{A} 
\frac{\partial \left( \xi^{-1} \ln \left\langle W_t \right\rangle \right)}{\partial \xi}
= - \frac{\partial \beta_s}{\partial \xi} C_{ss} 
  -  \frac{\partial \beta_t}{\partial \xi} C_{st} - \frac{\sigma a_t^2}{\xi} =0, \label{eq:dsigma1}\\
\frac{\partial (\sigma a_t^2)}{\partial \xi} &=& - \lim_{A \to \infty} \frac{1}{A} 
\frac{\partial \left( \xi^{-2} \ln \left\langle W_s \right\rangle \right)}{\partial \xi}
= - \frac{\partial \beta_s}{\partial \xi} C_{ts} 
  - \frac{\partial \beta_t}{\partial \xi} C_{tt} -2 \frac{\sigma a_t^2}{\xi} =0,
\label{eq:dsigma}
\end{eqnarray}
where, as mentioned in Sec.~\ref{sec:eos}, $a_t$ and $\xi$ are chosen as independent variables, and
$C_{xy}$ with $(x, y) = \{ s$ or $t \}$ is defined by 
\begin{eqnarray}
C_{xy} = \lim_{A \to \infty} \frac{1}{A \xi \langle W_x \rangle} 
\sum \left( \left\langle W_x P_y \right\rangle -\left\langle W_x \right\rangle
\left\langle P_y \right\rangle \right) 
\end{eqnarray}
with the sum taken over $y$-like plaquettes.
At $\xi=1$, $C_{ss}=C_{tt}$ and $C_{st}=C_{ts}$.
Then, the equations (\ref{eq:dsigma1}) and (\ref{eq:dsigma}) give
\begin{eqnarray}
(C_{ss} + C_{st}) \left(\frac{\partial \beta_s}{\partial \xi} 
+ \frac{\partial \beta_t}{\partial \xi}\right)
_{a_t : {\rm fixed},\, \xi = 1}  
= -3 \sigma a^2 . 
\label{eqn:csscst}
\end{eqnarray}
On the other hand, we also have
\begin{eqnarray}
\frac{d (\sigma a^2)}{d \beta} = 2 \sigma a \frac{d a}{d \beta}
= -C_{ss} -C_{st},
\end{eqnarray}
at $\xi=1$, which leads to the equation (\ref{eq:cubicsym}).

The other input to determine the anisotropy coefficients at $\xi=1$
can be obtained from the information about the phase transition point in the $(\beta_s, \beta_t)$ plane. 
The transition temperature $T_{c} = 1/[N_t a_t (\beta_s, \beta_t)]$ 
must be independent of the anisotropy of the lattice. 
Therefore, 
when we change the coupling constants, 
$ (\beta_s, \beta_t) \rightarrow 
(\beta_s + d \beta_s, 
\beta_t + d \beta_t) $ on a lattice with fixed $N_t$, 
along the transition curve, 
the lattice spacing in the temporal direction $a_t$ does not change:
\begin{eqnarray}
d a_{t }= \frac{\partial a_t}{\partial \beta_s} \,
d \beta_s + \frac{\partial a_t}{\partial \beta_t} \,
d \beta_t = 0.
\end{eqnarray}
Let us denote the slope of the transition curve at $\xi = 1$ as $r_t$,
\begin{eqnarray}
r_t =  
\frac{d \beta_s}{d \beta_t} = 
- \left(\frac{\partial a_t}{\partial \beta_t}\right)_{\xi = 1} 
\left/
  \left(\frac{\partial a_t}{\partial \beta_s}\right)_{\xi = 1}
\right. 
=  \left(\frac{\partial \beta_s}{\partial \xi}\right)_{\xi = 1} 
\left/ \left(\frac{\partial \beta_t}{\partial \xi}\right)_{\xi = 1}
\right. ,
\label{eq:rtr}
\end{eqnarray}
where we used an identity
\begin{eqnarray}
\left( \begin{array}{cc}
\frac{\partial \beta_s}{\partial a_t} &
\frac{\partial \beta_t}{\partial a_t} \\
\frac{\partial \beta_s}{\partial \xi} &
\frac{\partial \beta_t}{\partial \xi}
\end{array} \right) =
\left( \begin{array}{cc}
\frac{\partial a_t}{\partial \beta_s} &
\frac{\partial \xi}{\partial \beta_s} \\
\frac{\partial a_t}{\partial \beta_t} &
\frac{\partial \xi}{\partial \beta_t}
\end{array} \right)^{\! -1}
=
\frac{1}{D}
\left( \begin{array}{cc}
\frac{\partial \xi}{\partial \beta_t} &
-\frac{\partial \xi}{\partial \beta_s} \\
-\frac{\partial a_t}{\partial \beta_t} &
\frac{\partial a_t}{\partial \beta_s}
\end{array} \right),
\end{eqnarray}
with
$
D=
\frac{\partial \xi}{\partial \beta_t}
\frac{\partial a_t}{\partial \beta_s}
-\frac{\partial \xi}{\partial \beta_s}
\frac{\partial a_t}{\partial \beta_t}
$.
From Eqs.~(\ref{eq:cubicsym}) and (\ref{eq:rtr}), the derivatives of 
$\beta_s$ and $\beta_t$ with respect to $\xi$ are expressed as 
\begin{eqnarray}
\left(\frac{\partial \beta_s}{\partial \xi}\right)_{a_t : {\rm fixed},\, \xi = 1} 
&=& \frac{3 r_t}{2 ( 1 + r_t )} \,
a \frac{d \beta}{d a}, \nonumber \\
 \left(\frac{\partial \beta_t}{\partial \xi}\right)_{a_t : {\rm fixed},\, \xi = 1} 
&=& \frac{3}{2 ( 1 + r_t )} \, a \frac{d \beta}{d a}. 
\label{eq:anisocoeff}
\end{eqnarray}
Using the slope $r_t$ and the beta function, 
the conventional combinations $\epsilon-3p$ and $\epsilon+p$
are given by 
\begin{eqnarray}
(\epsilon-3p)/T^4 &=& 
- 3 N_t^4 \, a \frac{d\beta}{da} \,
\{\langle P_s \rangle + \langle P_t \rangle\ - 2\langle P \rangle_0\}, 
\label{eq:e3p} \\
(\epsilon+p)/T^4 &=& 
3 N_t^4 \, a \frac{d\beta}{da} \, 
\frac{r_t-1}{r_t+1} \,
\{\langle P_s \rangle - \langle P_t \rangle\}.
\label{eq:emp}
\end{eqnarray}

Moreover, introducing the notation
$\gamma = \sqrt{\beta_t/\beta_s}$, 
we obtain
\begin{eqnarray}
\left(\frac{\partial\gamma}{\partial\xi}\right)_{\! a_t:{\rm fixed},\,\xi=1}
= \left(\frac{\partial\gamma}{\partial\xi}\right)_{\! a_s:{\rm fixed},\,\xi=1}
= \frac{3}{4\beta} \, \frac{1-r_t}{1+r_t} \, a\frac{d\beta}{da}.
\label{eq:dgamdxi}
\end{eqnarray}

Finally, the customarily used forms for the anisotropy coefficients 
(Karsch coefficients) \cite{karsch} are given by 
\begin{eqnarray}
c_s 
&=& \left(\frac{\partial g_s^{-2} }
    {\partial \xi}\right)_{\! a_s: {\rm fixed},\, \xi = 1} 
= \frac{1}{2N_{c}} \left\{ \beta + \frac{r_t - 2}{2 ( 1 + r_t)} \,
    a \frac{d \beta}{d a} \right\}, \nonumber \\ 
c_t 
&=& \left(\frac{\partial g_t^{-2} }
    {\partial \xi}\right)_{\! a_s: {\rm fixed},\, \xi = 1} 
= \frac{1}{2N_{c}} \left\{ - \beta 
    + \frac{1 - 2 r_t}{2 ( 1 + r_t)} \, 
    a \frac{d \beta}{d a} \right\},
\label{kc}
\end{eqnarray}
where 
$\beta_s = 2N_{c} g_s^{-2} \xi^{-1}$ and 
$\beta_t = 2N_{c} g_t^{-2} \xi$.
Therefore, when the value for the beta function is available, 
we can determine these anisotropy coefficients by measuring $r_t$ 
from the finite temperature transition line in 
the $(\beta_s, \beta_t)$ plane.

\subsection{Slope of the transition line}
\label{sec:transition}

In order to determine the transition line in the coupling 
parameter space, we calculate the rotated Polyakov loop 
\begin{eqnarray}
\Omega = z \, \frac{1}{N_s^3} \sum_{\vec{x}} \frac{1}{N_{c}} 
{\rm Tr} \prod_{t=1}^{N_t} U_4( \vec{x},t )
\end{eqnarray}
as a function of $(\beta_s, \beta_t)$, where $z$ is a $Z(N_c)$ phase 
factor ($z^{N_c} = 1$) such that $\arg(\Omega) \in (-\pi/N_c,\pi/N_c]$.
Thus, $\Omega$ is a complex number.
We define the transition point as the peak position of 
the susceptibility 
\begin{eqnarray}
\chi_{\Omega} = N_s^3\left(\langle \Omega^2 \rangle - \langle \Omega \rangle^2\right) .
\end{eqnarray}

In Sec.~\ref{sec:reweighting}, we investigate the coupling parameter dependence of $\chi_{\Omega}$ on the $(\beta_s,\beta_t)$ plane 
by applying the multi-point reweighting method. 
The reweighting method enables us to compute the anisotropy coefficients 
directly from simulations just at $\xi = 1$ 
without introducing an interpolation ansatz. 
Therefore, we can use data of previous high statistic simulations on isotropic lattices. 
In particular, this is a great advantage for a computation of the latent heat 
because high statistic simulations are required for a precise calculation of 
the plaquette gap between two phases at the transition point. 
The high statistic data can be used also for determination of the phase transition line and  
the anisotropy coefficients.

As we see in the next section,  $\chi_{\Omega}$ forms a ridge approximately in the $\gamma$ direction on the $(\beta_s,\beta_t)$ plane. 
This is due to the fact that the transition temperature is independent of the anisotropy $\xi$.
Therefore, to determine peak positions, it is convenient to introduce  $\bar\beta = \sqrt{\beta_s \beta_t}$ which is perpendicular to the $\gamma$ direction at $\gamma\approx1$.
The slope in  the $(\beta_s,\beta_t)$ plane is now given by
\begin{eqnarray}
\frac{d\beta_s}{d\beta_t} 
= \frac{d ( \bar\beta / \gamma )}{d ( \bar\beta \gamma )}
= \frac1{\gamma^2} \, 
\frac{\gamma (d\bar\beta/d\gamma) - \bar\beta }
{\gamma (d\bar\beta/d\gamma) + \bar\beta } \, .
\end{eqnarray}
Denoting the transition point for given $\gamma$ as $\bar\beta_c(\gamma)$, and 
fitting $\bar\beta_c(\gamma)$ with a polynomial 
\begin{equation}
 \bar\beta_{c} (\gamma) = \sum_{n = 0}^{n_{\rm max}} f_n \, (\gamma - 1)^{n},
\label{eq:polynomial}
\end{equation}
with $f_n$ the fitting parameters, 
the slope $ r_t$ of the transition line at $\xi=1$ ($\gamma=1$) is given by 
\begin{eqnarray}
r_t = \frac{ f_1 - \beta_{c}} 
{ f_1 + \beta_{c}} \, .
\end{eqnarray}
We confirm that the results are completely stable under a variation 
of $n_{\rm max}$ and the fitting range of $\gamma$.

\subsection{Condition for vanishing pressure gap}
\label{sec:gap}

From Eq.~(\ref{eq:dprs}), when the pressure gap $\Delta p$ vanishes, 
\begin{eqnarray}
\left. \frac{\partial \beta_s}{\partial \xi} \right/ 
\frac{\partial \beta_t}{\partial \xi} 
= -\frac{\langle P_t \rangle_{\rm hot}- \langle P_t \rangle_{\rm cold}}
{\langle P_s \rangle_{\rm hot}- \langle P_s \rangle_{\rm cold}} 
\end{eqnarray}
should be satisfied.
The left hand side is related to the slope of the transition line $r_t$ by Eq.~(\ref{eq:rtr}).
Hence, the condition for $\Delta p=0$ reads \cite{ejiri03}
\begin{eqnarray}
\frac{\langle P_t \rangle_{\rm hot}- \langle P_t \rangle_{\rm cold}}
{\langle P_s \rangle_{\rm hot}- \langle P_s \rangle_{\rm cold}} = -r_t.
\label{eq:dp0cond}
\end{eqnarray}
In the next section, we test if this relation holds.

\subsection{Multi-point reweighting method}
\label{sec:reweighting}

To find the transition line in the $(\beta_s, \beta_t)$ plane,
we need the expectation values of an order parameter and its susceptibility as continuous functions of $(\beta_s, \beta_t)$. 
In lattice simulations, the reweighting method \cite{MDS67,FS89} is useful in varying coupling parameters continuously.
Around a first order phase transition point, however, large fluctuation of the reweighting factor due to the flip-flop between two phases can make the applicability range of a reweighting method very small.
Here, it is noted in Ref.~\cite{ejiri03} that,
when we shift $\beta_s$ and $\beta_t$ around the first order phase transition point, 
the leading fluctuations of the reweighting factor in $\Delta\beta_s$ and $\Delta\beta_t$ due to the flip-flop cancel out with each other if $\Delta\beta_s/\Delta\beta_t = r_t$ satisfies Eq.~(\ref{eq:dp0cond}). 
This means that the reweighting method is applicable for the determination of the transition line.

To further extend the applicability range in the coupling parameter space, 
we adopt the multi-point reweighting method \cite{FS89,iwami15}: 
Let us define the histogram for a set of observables 
$X = (X_1,X_2,\cdots)$ as 
\begin{eqnarray}
w(X; \beta) 
&=& \int {\cal D} U \, \prod_i \delta(X_i - \hat{X}_i) \ e^{- S} .
\label{eq:dist}
\end{eqnarray}
where $\hat{X} = (\hat{X}_1,\hat{X}_2,\cdots)$ is the operators for $X$. 
For simplicity, we denote $(\beta_s, \beta_t)$ as $\vec{\beta}$ and use the notation 
$(\vec{\beta} \cdot \vec{P}) = \beta_s P_s + \beta_t P_t$. 
The action is then $S= -3N_{\rm site} (\vec{\beta} \cdot \vec{P})$.
Using $w(X; \vec{\beta})$, the partition function is given by
$ Z(\vec{\beta})  = \int w(X; \vec{\beta}) \, dX $
with $dX = \prod_i dX_i$, 
and the probability distribution function of $X$ is given by $Z^{-1} w(X; \vec{\beta})$. 
The expectation value of an operator ${\cal O} [\hat{X}]$ which is written in terms of $\hat{X}$ is calculated by 
\begin{eqnarray}
\langle {\cal O }[\hat{X}] \rangle_{\vec{\beta}} = \frac{1}{Z(\vec{\beta}) } 
\int\! {\cal O} [X] \, w(X; \vec{\beta}) \, dX .
\label{eq:expop}
\end{eqnarray}

To obtain $w(X; \vec{\beta})$ which is reliable in a wide range of $X$, we make use of the reweighing formulas to combine data obtained at different simulation points \cite{FS89}. 
We combine a set of $N_{\rm sp}$ simulations performed at $\vec{\beta}_i=(\beta_{si}, \beta_{ti})$ with the number of configurations $N_i$ where $i=1, \cdots , N_{\rm sp}$.
Here, we choose $P_s$ and $P_t$ as two observables of  $X$ and redefine $X$ as the set of observables other than $\vec{P} = (P_s,P_t)$.
From the definition Eq.~(\ref{eq:dist}), the probability distribution function at $\vec{\beta}_i$ is related to that at $\vec{\beta}=(\beta_s, \beta_t)$ as  
\begin{eqnarray}
Z^{-1}(\vec{\beta}_i) \,w(\vec{P},X; \vec{\beta}_i) 
= Z^{-1}(\vec{\beta}_i) \, e^{3N_{\rm site} ((\vec{\beta}_i-\vec{\beta}) \cdot \vec{P})} 
 \,w(\vec{P}, X; \vec{\beta}) .
\end{eqnarray} 
Summing up these probability distribution functions with the weight $N_i$, 
\begin{eqnarray}
\sum_{i=1}^{N_{\rm sp}} N_i \, Z^{-1}(\vec{\beta}_i) \, w(\vec{P},X; \vec{\beta}_i) 
= e^{-3N_{\rm site} (\vec{\beta} \cdot \vec{P})} 
\sum_{i=1}^{N_{\rm sp}} N_i \, Z^{-1}(\vec{\beta}_i) \, e^{3N_{\rm site} (\vec{\beta}_i \cdot \vec{P})}  \, w(\vec{P}, X; \vec{\beta}), 
\label{eq:sum1}
\end{eqnarray}
we obtain 
\begin{eqnarray}
w(\vec{P}, X; \vec{\beta})= G(\vec{P}; \vec{\beta}, \vec{B}) \,
\sum_{i=1}^{N_{\rm sp}} N_i \, Z^{-1}(\vec{\beta}_i) \, w(\vec{P},X; \vec{\beta}_i) 
\end{eqnarray}
with the simulation points $\vec{B}=(\vec{\beta}_1,\cdots,\vec{\beta}_{N_{\rm sp}})$ and 
\begin{eqnarray}
G(\vec{P}; \vec{\beta}, \vec{B})=\frac{ e^{3N_{\rm site} (\vec{\beta} \cdot \vec{P})}}{
\sum_{i=1}^{N_{\rm sp}} N_i \, e^{3N_{\rm site} (\vec{\beta}_i \cdot \vec{P})} Z^{-1}(\vec{\beta}_i)} .
\end{eqnarray}
Note that the left-hand side of Eq.~(\ref{eq:sum1}) gives a naive histogram using all the configurations disregarding the difference in the simulation parameter.
The histogram $w(\vec{P}, X; \vec{\beta})$ at $\vec{\beta}$ is given by 
multiplying $G(\vec{P}; \vec{\beta}, \vec{B})$ to this naive histogram.

The partition function is given by
\begin{eqnarray}
Z(\vec{\beta})= \sum_{i=1}^{N_{\rm sp}} N_i \int G(\vec{P}; \vec{\beta}, \vec{B}) \, Z^{-1}(\vec{\beta}_i) \, w(\vec{P},X; \vec{\beta}_i) \, dP \,dX 
=\sum_{i=1}^{N_{\rm sp}} N_i \left\langle G(\vec{\hat{P}}; \vec{\beta}, \vec{B}) \right\rangle_{(\vec{\beta}_i)}.
\end{eqnarray}
The right-hand side is just the naive sum of $G(\vec{\hat{P}}; \vec{\beta}, \vec{B})$ observed on all the configurations.
The partition function at $\vec{\beta}_i$ can be determined, up to an overall factor, by the consistency relations, 
\begin{eqnarray}
Z(\vec{\beta}_i) 
=\sum_{k=1}^{N_{\rm sp}} N_k \left\langle G(\vec{\hat{P}}; \vec{\beta}_i, \vec{B}) \right\rangle_{(\vec{\beta}_k)}
=\sum_{k=1}^{N_{\rm sp}} N_k \left\langle 
\frac{e^{3N_{\rm site} (\vec{\beta}_i \cdot \vec{\hat{P}})}}{
\sum_{j=1}^{N_{\rm sp}} N_j e^{3N_{\rm site} (\vec{\beta}_j \cdot \vec{\hat{P}})} Z^{-1}(\vec{\beta}_j)} \right\rangle_{(\vec{\beta}_k)}
\end{eqnarray}
for $i=1,\cdots,N_{\rm sp}$. 
Denoting $f_i=-\ln Z(\vec{\beta}_i)$, these equations can be rewritten by
\begin{eqnarray}
1 = \sum_{k=1}^{N_{\rm sp}} N_k \left\langle
\frac{1}{ \sum_{j=1}^{N_{\rm sp}} N_j \exp[ 3N_{\rm site} ((\vec{\beta}_j- \vec{\beta}_i) \cdot \vec{\hat{P}}) 
- f_i +f_j]} \right\rangle_{(\vec\beta_k)},
\hspace{5mm}
i=1,\cdots,N_{\rm sp}.
\label{eq:consis}
\end{eqnarray}
Starting from appropriate initial values of $f_i$, we solve these equations numerically by an iterative method. 
Note that, in these calculations, one of the $f_i$'s must be fixed to remove the ambiguity corresponding to the undetermined overall factor.

Then, the expectation value of an operator ${\cal O}[\vec{\hat{P}},\hat{X}]$ at $\vec{\beta}$, 
Eq.~(\ref{eq:expop}), can be evaluated as
\begin{eqnarray}
\langle {\cal O} [\vec{\hat{P}},\hat{X}] \rangle_{(\beta)} 
= \frac{1}{Z(\vec{\beta})} \sum_{i=1}^{N_{\rm sp}} N_i \left\langle {\cal O} [\vec{\hat{P}},\hat{X}] \, G(\vec{\hat{P}}; \vec{\beta}, \vec{B}) \right\rangle_{(\vec{\beta}_i)}.
\label{eq:multibeta}
\end{eqnarray}
Again, 
$\sum_{i=1}^{N_{\rm sp}} N_i \left\langle {\cal O} G \right\rangle_{(\beta_i)}$ 
in the right-hand side is just the naive sum of ${\cal O} G$ over all the configurations disregarding the difference in the simulation point.

\section{Results}
\label{sec:results}

\subsection{Simulation parameters}
\label{sec:parameter}

\begin{table}
\caption{Summary of the simulation setup. The lattice size is $N_s^3 \times N_t$. 
$N_{\rm conf.}$ is the number of configurations after thermalization.}
\label{tab:parameter}
\begin{tabular}{cclr}
\hline
\hline
$N_s$ & $N_t$ & $\beta$ & $N_{\rm conf.}$ \\
\hline
48&6 &5.89379&201200\\
\hline
64&6 &5.893  &30000\\
64&6 &5.89379&150000\\
64&6 &5.894  &215000\\
64&6 &5.895  &47000\\
\hline
48&8 &6.056  &200000\\
48&8 &6.058  &200000\\
48&8 &6.06   &200000\\
48&8 &6.062  &200000\\
48&8 &6.065  &220000\\
48&8 &6.067  &200000\\
\hline
64&8 &6.0585 &95000\\
64&8 &6.061  &2060000\\
64&8 &6.063  &300000\\
64&8 &6.065  &510000\\
64&8 &6.068  &1620000\\
\hline
64&12&6.3335 &324000\\
64&12&6.335  &290000\\
64&12&6.3375 &10000\\
\hline
96&12&6.332  &45000\\
96&12&6.334  &474000\\
96&12&6.335  &534000\\
96&12&6.336  &336000\\
96&12&6.338  &169000\\
\hline
\hline
\end{tabular}
\end{table}

On isotopic lattices, 
i.e. $\beta_s = \beta_t = \beta$, $\xi = a_s/a_t =1$, 
we perform simulations of the SU(3) gauge theory at several $\beta$ points around the deconfining phase transition point. 
The lattice sizes for temporal direction are $N_t=6, 8$ and $12$ with two different volumes for each $N_t$.
Our simulation parameters are summarized in Table~\ref{tab:parameter}.
The configurations are generated by a pseudo heat bath algorithm followed by 5 over-relaxation sweeps.
The Polyakov loop and the plaquettes are measured every iteration.
Data are taken at one to five $\beta$ values for each $(N_s,N_t)$ 
and are combined using the multi-point reweighting method discussed in Sec.~\ref{sec:reweighting}.

The number of flip-flops between the hot and cold phases should not be small to obtain statistically reliable results near the first order transition point.
We count the number of flip-flops during the Monte-Carlo 
simulations by the method we explain in Sec.~\ref{sec:phase}.
These are 16 times at $\beta= 6.335$ on the $96^3 \times 12$ lattice, 
and 116 times at $\beta= 6.061$ on the $64^3 \times 8$ lattice, for example.
Flip-flops happen more frequently when the aspect ratio $N_s/N_t$ gets smaller.
Thus, our numbers of flip-flops would be sufficient.
The statistical errors are estimated by the jack-knife method. 
The bin size is adopted to be 1000, which is much smaller than the typical size of the interval of flip-flops.
The errors are saturated with this bin size.

For continuum and large volume extrapolations, we include the data obtained on 
$36^2 \times 48 \times 6$ lattice by the QCDPAX Collaboration \cite{QCDPAX}. 
The aspect ratios of this lattice, $V^{1/3}/N_t = 6.6$ with $V$ the spatial volume of the lattice, is within the range of our aspect ratios, $N_s/N_t = 5.3$--10.7, while those of other $N_t=6$ lattices of Ref.~\cite{QCDPAX} are less than 4.

\subsection{Transition line and anisotropy coefficients}
\label{sec:aniso}

\begin{figure}[tb]
\centering
\includegraphics[width=90mm]{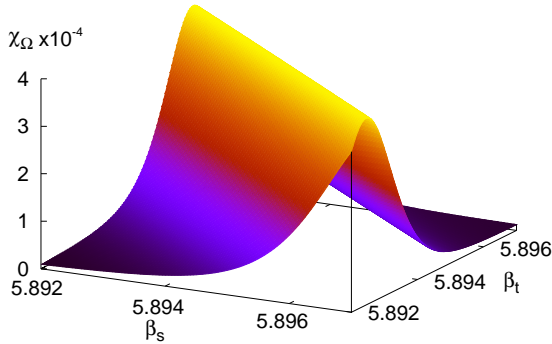}
\caption{Polyakov loop susceptibility $\chi_{\Omega}$ as a function of 
$(\beta_s, \beta_t)$ obtained on a $64^3 \times 6$ lattice.
}
\label{fig:plsus}
\end{figure}

\begin{figure}[tb]
\centering
\includegraphics[width=75mm]{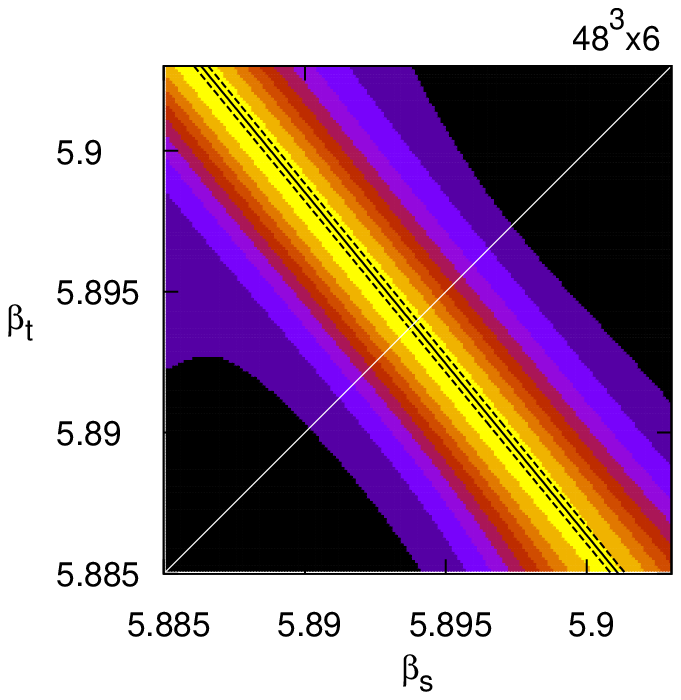}
\hspace{2mm}
\includegraphics[width=75mm]{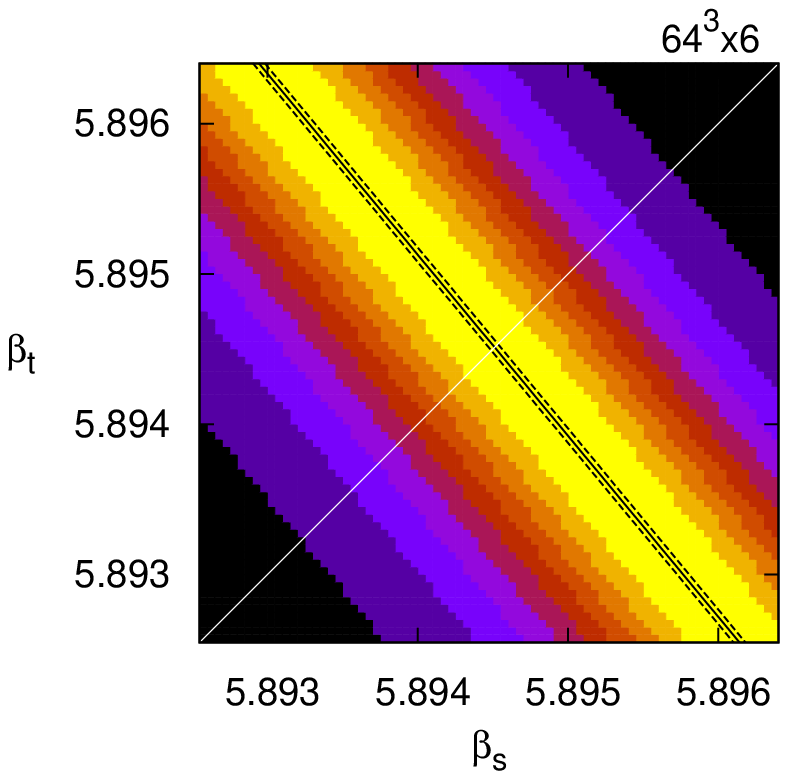} \\
\vspace{-3mm}
\includegraphics[width=75mm]{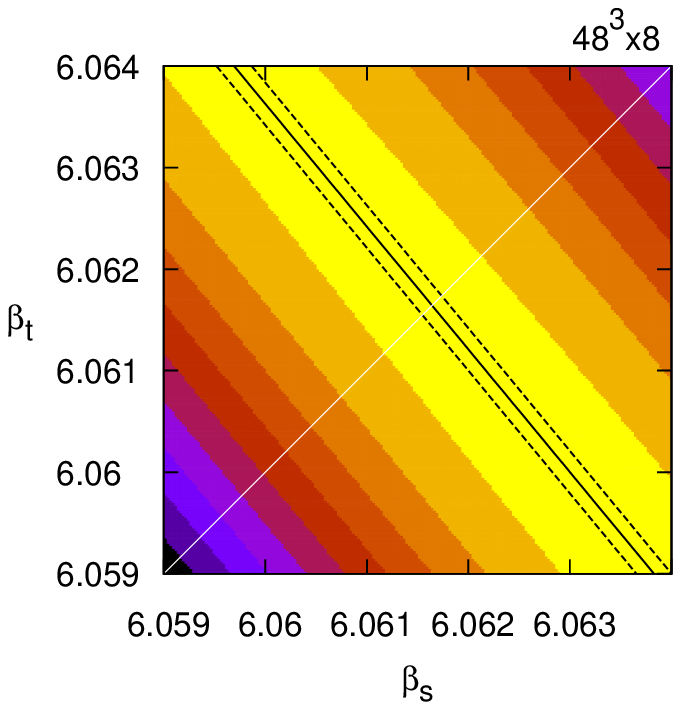}
\hspace{2mm}
\includegraphics[width=75mm]{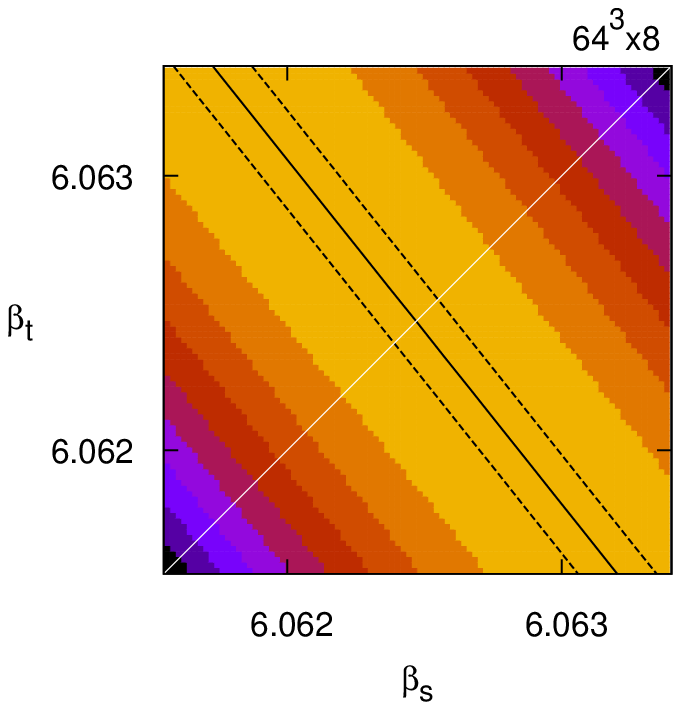} \\
\vspace{-3mm}
\includegraphics[width=75mm]{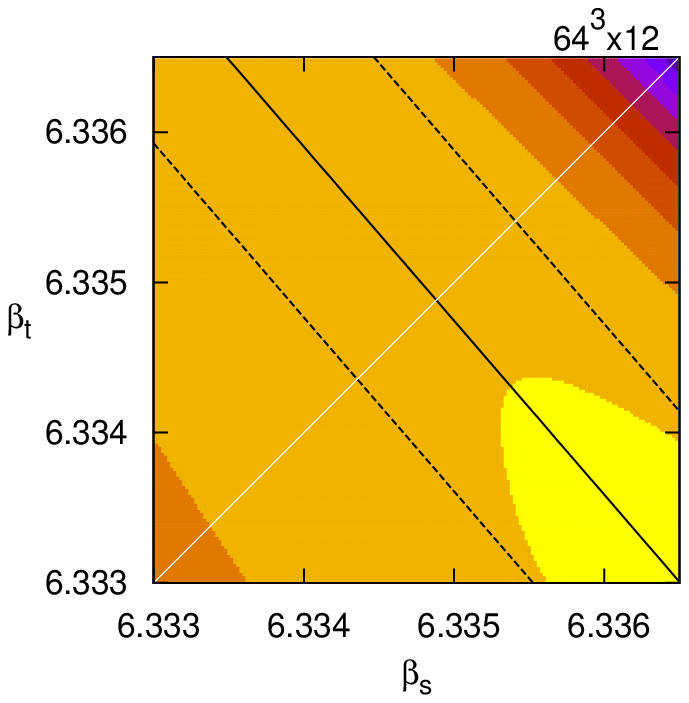}
\hspace{2mm}
\includegraphics[width=75mm]{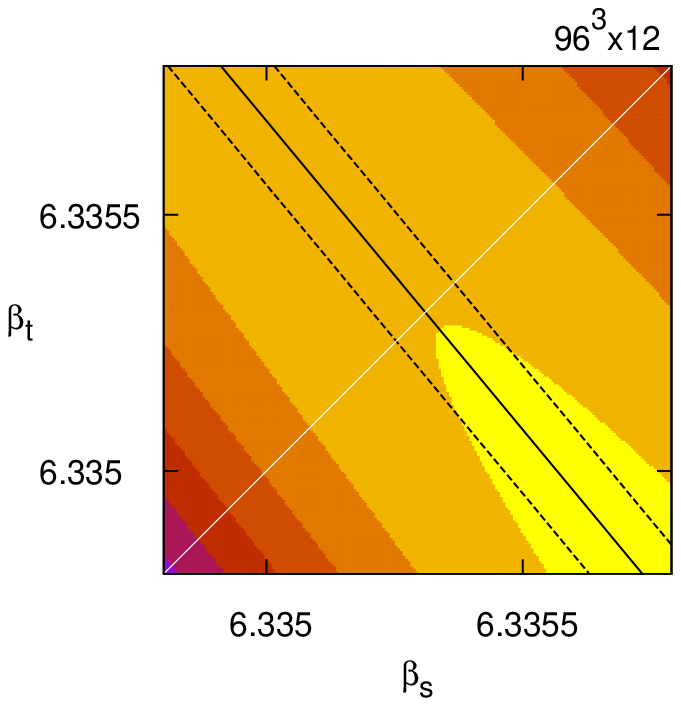}
\vspace{-2mm}
\caption{Contour plots of the Polyakov loop susceptibility 
as a function of $(\beta_s, \beta_t)$.
The lattice sizes are 
$48^3 \times 6$ (top left), $64^3 \times 6$ (top right), 
$48^3 \times 8$ (middle left), $64^3 \times 8$ (middle right), 
$64^3 \times 12$ (bottom left) and $96^3 \times 12$ (bottom right), 
respectively.
The solid lines indicate the phase transition line on each figure and the 
dashed lines are the upper and lower bounds of the error.
}
\label{fig:suscp}
\end{figure}

\begin{figure}[tb]
\centering
\includegraphics[width=95mm]{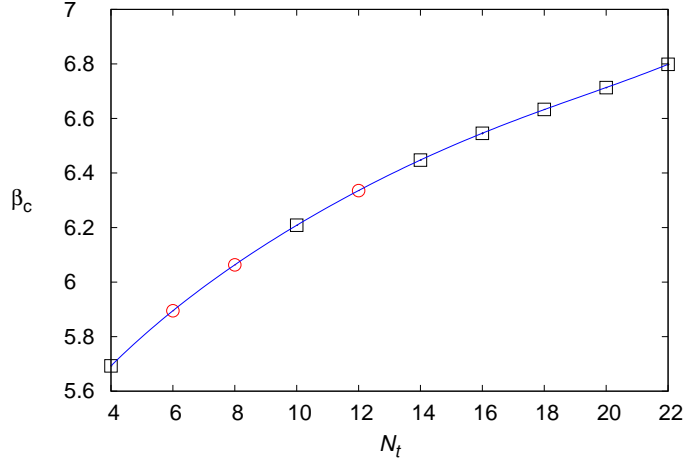}
\vspace{-2mm}
\caption{The transition point $\beta_c$ as a function of $N_t$.
Our results at $N_t=6$, 8 an 12 are shown by circular symbols. 
The data at $N_t=4$, 10, and $14$ -- $22$ (squares) are obtained in Ref.~\cite{francis15}.
The errors are much smaller than the symbols.
The blue curve is the result of fitting discussed in the text.
}
\label{fig:tc}
\end{figure}

\begin{figure}[tb]
\centering
\includegraphics[width=80mm]{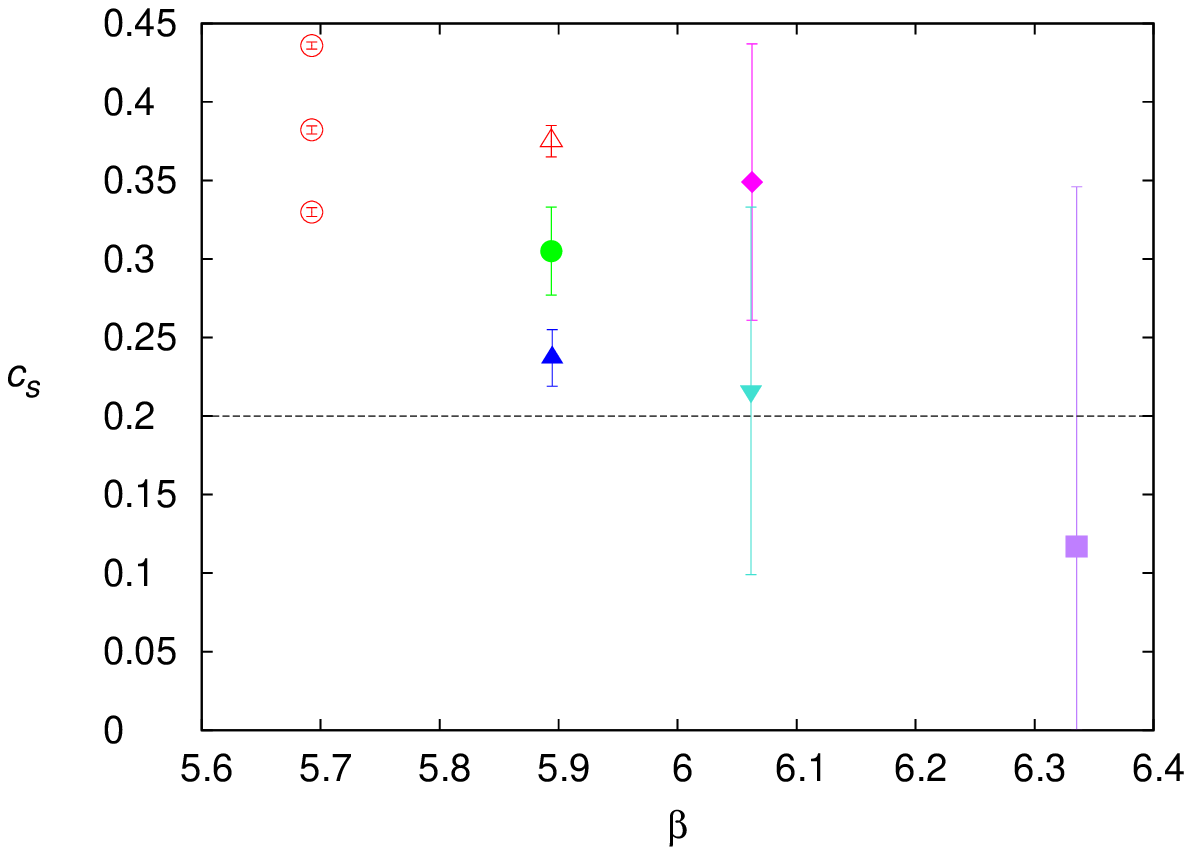}
\includegraphics[width=80mm]{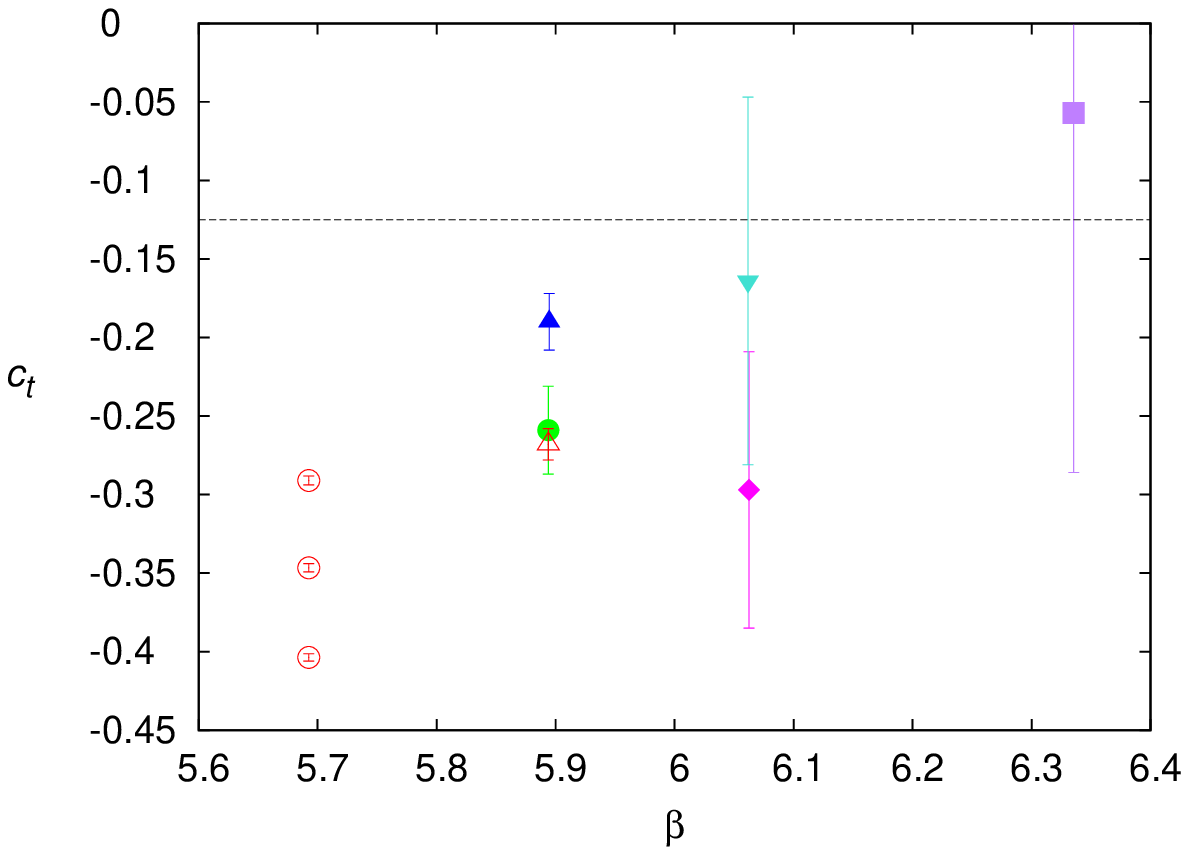}
\caption{Results of the Karsch coefficients $c_s$ (left) and $c_t$ (right) 
as functions of $\beta$. The dashed lines are the perturbative values.
The three open circles at $\beta = 5.69245$ are the results of Ref.~\cite{ejiri98} obtained on a $24^2\times36\times4$ lattice with three different beta functions (beta functions defined by $\beta_c$ \cite{boyd}, string tension  \cite{edwards}, and a block spin transformation \cite{taro}, from bottom to top for $c_s$ and in reverse order for $c_t$).
}
\label{fig:cst}
\end{figure}

\begin{table}[tb]
\caption{Results of $\beta_c$ and the slope $r_t$ at $\xi=1$ on each lattice, together with the ratio 
$(\left\langle P_t \right\rangle_{\rm hot} - \left\langle P_t \right\rangle_{\rm cold})
/(\left\langle P_s \right\rangle_{\rm hot} - \left\langle P_s \right\rangle_{\rm cold})$.
The column ``$\gamma$-range'' is for the range of $\gamma$
used in the fit for the slope. 
We compare $r_t$ and  
$(\left\langle P_t \right\rangle_{\rm hot} - \left\langle P_t \right\rangle_{\rm cold})
/(\left\langle P_s \right\rangle_{\rm hot} - \left\langle P_s \right\rangle_{\rm cold})$
in Sec.~\ref{sec:phase}. 
}
\label{tab:rt}
\begin{tabular}{llcll}
\hline
\hline
lattice       & $\beta _c(N_t,V)$ & $\gamma$-range & $r_t$ &
$\frac{\left\langle P_t \right\rangle_{\rm hot} -\left\langle P_t \right\rangle_{\rm cold}}{\left\langle P_s \right\rangle_{\rm hot} - \left\langle P_s \right\rangle_{\rm cold}}$ \\
\hline
$48^3\times6$ &5.89383(24) &0.999--1.001  &-1.2020(39)&1.216(50)\\
$64^3\times6$ &5.894512(40)&0.999--1.001  &-1.2022(52)&1.2053(38)\\
$48^3\times8$ &6.06160(18) &0.9998--1.0002&-1.209(33) &1.204(14)\\
$64^3\times8$ &6.06247(14) &0.9998--1.0002&-1.255(37) &1.2344(66)\\
$64^3\times12$&6.3349(11)  &0.9998--1.0002&-1.16(61)  &1.327(84)\\
$96^3\times12$&6.33533(11) &0.9999--1.0001&-1.204(53) &1.283(53)\\
\hline
\hline
\end{tabular}
\end{table}

\begin{table}[tb]
\caption{Beta function and anisotropy coefficients at $\xi=1$ and at the transition point on each lattice. 
}
\label{tab:cst}
\begin{tabular}{lllllll}
\hline
\hline
lattice & $\beta_c(N_t,V)$ & $a(d\beta/da)$&$\partial \gamma / \partial \xi$&$c_s$&$c_t$\\\hline
$36^2\times48\times6$&5.89379(34) &-0.5483(8) &0.704(10) &0.375(10) &-0.268(10)\\
$48^3\times6$        &5.89383(24)   &-0.5484(8) &0.761(15) &0.303(28) &-0.257(28)\\
$64^3\times6$        &5.894512(40) &-0.5486(8) &0.760(20) &0.258(17) &-0.213(17)\\
$48^3\times8$        &6.06160(18)   &-0.6210(8) &0.813(127)&0.215(117)&-0.163(117)\\
$64^3\times8$        &6.06247(14)   &-0.6214(8) &0.681(99) &0.349(88) &-0.297(88)\\
$64^3\times12$      &6.3349(11)     &-0.7164(12)&1.16(450) &-0.16(414)&0.22(414)\\
$96^3\times12$      &6.33532(11)   &-0.7165(12)&0.915(239)&0.119(228)&-0.060(228)\\
\hline
\hline
\end{tabular}
\end{table}

The result of the Polyakov loop susceptibility $\chi_{\Omega}$ on the $64^3 \times 6$ lattice
is plotted in Fig.~\ref{fig:plsus} as a function of $(\beta_s, \beta_t)$.
Because the transition is of first order for the SU(3) gauge theory, the peak of $\chi_{\Omega}$
is quite clear with our large spatial volumes. 
Contour plots of $\chi_{\Omega}$ are collected in Fig.~\ref{fig:suscp}.
A brighter color means a larger $\chi_{\Omega}$.
The phase transition line $\bar\beta_c(\gamma)$ is defined as the peak position of the susceptibility 
for each $\gamma$. 
The results of $\bar\beta_c(\gamma)$ are shown by solid lines in Fig.~\ref{fig:suscp}, 
with the dashed lines their jackknife errors.

We now calculate the slope $r_t$ by the the method discussed in Sec.~\ref{sec:transition}.
We choose the fit ranges of $\gamma$ for the polynomial fit Eq.~(\ref{eq:polynomial}) such that the statistical error of the susceptibility is sufficiently small and the transition line is approximately straight.
The fit ranges are summarized in Table~\ref{tab:rt}.
We confirm that the fit range dependence is small in the results.
We also study the dependence on the largest order of the polynomial in Eq.~(\ref{eq:polynomial}) by varying $n_{\rm max}=1$--7.
We find that the fits work well and stable for $n_{\rm max}\ge3$.
The differences in $r_t$ between $n_{\rm max}=3$ and $4$ are less than 0.5\% and
are much smaller than the statistical errors.
We thus adopt $n_{\rm max}=3$.
Our results of the transition point $\beta_c$ and the slope $r_t$ at $\xi=1$ are summarized in Table~\ref{tab:rt}.
From this Table, we find no clear spatial volume dependence in $r_t$. 

Unlike the case of $r_t$, we do expect that $\beta_c$ has the spatial volume dependence following the finite size scaling theory.
Accordingly, $\beta_c$ in Table~\ref{tab:rt} shows some spatial volume dependence.
To calculate the beta function, 
we thus extrapolate the results of $\beta_c$ to the infinite spatial volume limit adopting the finite size scaling relation of first order phase transition,
$
\beta_c(N_t,V) - \beta_c(N_t,\infty) \propto 1/V
$,
where $V$ is the spatial volume of the lattice.
For the fit at $N_t=6$, we include the data obtained on the $36^2 \times 48 \times 6$ lattice of the QCDPAX Collaboration \cite{QCDPAX}.
We obtain $\beta_c(N_t,\infty) = 5.894512(40)$,  6.06307(28) and 6.33552(47) at $N_t=6$, 8, and 12, respectively.

The results of $\beta_c(N_t,\infty)$ as a function of $N_t$ are summarized in Fig.~\ref{fig:tc}, together with the data at $N_t=4$, 10, and $14$--$22$ reported in Ref.~\cite{francis15}.
The difference is invisible in the Figure when we replace our $\beta_c(N_t,\infty)$ with $\beta_c(N_t,V)$ on our finite lattices.
We now calculate the non-perturbative beta function using the fact that the lattice spacing is $a=1/(N_t T_c)$ at the transition point $\beta_c(N_t,\infty)$.
We fit the data of Fig.~\ref{fig:tc} by 
\begin{eqnarray}
\beta_c (N_t,\infty) = \sum_{n=0}^{n_{\rm max}} b_n \, N_t ^{\, n}
\end{eqnarray}
with $b_n$ the fit parameters.
The fit result is shown in Fig.~\ref{fig:tc} with a blue curve. 
We adopt $n_{\rm max}=5$ with which we obtain $\chi^2/{\rm dof} = 1.3$. 
We confirm that the $n_{\rm max}$ dependence is quite small in the final results:
The differences in $a (d \beta /da)$ computed with $n_{\rm max}=5$ and $6$ 
are about $0.2\%$ at $N_t=6$ and 8, and is about $0.6\%$ at $N_t=12$,  
which are much smaller than the statistical error of $r_t$.
From the resulting fit function, we compute the beta function as $a (d \beta /da) = -N_t (d \beta_c/d N_t)$.
We obtain $a (d \beta /da) = -0.5488(8)$, $-0.6217(8)$ and $-0.7166(26)$ at $\beta_c(N_t,\infty)$ for $N_t=6$, 8 and 12, respectively.
The results of the beta function at $\beta_c(N_t,V)$ are given in Table~\ref{tab:cst}.
We note that the statistical errors in $a (d \beta /da)$ are very small.

Combining the results of $a (d \beta /da)$ and $r_t$ on each lattice, we compute the anisotropy coefficients using Eqs.~(\ref{eq:dgamdxi}) and (\ref{kc}), 
as summarized in Table~\ref{tab:cst}.
In this Table, we also list the beta function and anisotropy coefficients on the $36^2 \times 48 \times 6$ lattice \cite{QCDPAX}, for later use.
For this lattice, we combine our $a (d \beta /da)$ with the results of $r_t$ obtained in Ref.~\cite{ejiri98} on this lattice.

The results of the Karsch coefficients $c_s$ and $c_t$ are plotted in Fig.~\ref{fig:cst} as functions of $\beta$.
The dashed lines represent the perturbative values \cite{karsch}.
Results on the $64^3\times12$ lattice are omitted due to the large errors. 

For comparison, we also show in Fig.~\ref{fig:cst} the results of non-perturbative anisotropy coefficients at $\beta_c = 5.69245$ of $N_t=4$ obtained on a $24^2\times36\times4$ lattice \cite{ejiri98}.  
As noted in Ref.~\cite{ejiri98}, the uncertainty due to the choice of the beta function is not small at $\beta_c$ of $N_t=4$: 
Among three beta functions defined by $\beta_c$ \cite{boyd}, string tension \cite{edwards}, and a block spin transformation \cite{taro}, the difference is as large as 19\% at $N_t=4$, 
while it is about $3.5\%$ at $N_t=6$.
To get an idea about this uncertainty at $N_t=4$, we show all the results using these three beta functions by the three open circles in Fig.~\ref{fig:cst}.

We find that the clear deviations from the perturbative values at $N_t=4$ and 6 decrease as $N_t$ is increased ($\beta$ is increased).
The overall $\beta$-dependence of $c_s$ and $c_t$ suggests that the perturbative values may be reproduced around $N_t=12$ ($\beta \sim 6.3$) though the errors are large.
\footnote{
In Ref.~\cite{engels00}, the authors attempted to estimate $c_s$ and $c_t$ by imposing a condition that the result of the equation of state by the integral method should be reproduced, and suggested deviation of $c_s$ and $c_t$ from the perturbation theory also at around $\beta_c$ of $N_t=12$. 
Accordingly, our direct calculation of $c_s$ and $c_t$ leads to the equation of state slightly different from the results of the integral method at $N_t=12$, though both methods lead to vanishing pressure gap.
Here, we note that, on finite lattices, the results of the integral method and the derivative method do not agree with each other even in the high temperature limit: 
The derivative method leads to 6.8\% larger energy density than that by the integral method on $96^3 \times 12$ lattice and 7.2\% larger on $64^3 \times 8$ lattice, for example. 
This may explain the difference in the values of $c_s$ and $c_t$.
}

\subsection{Phase separation at the first order transition point}
\label{sec:phase}

\begin{figure}[tb]
\centering
\includegraphics[width=80mm]{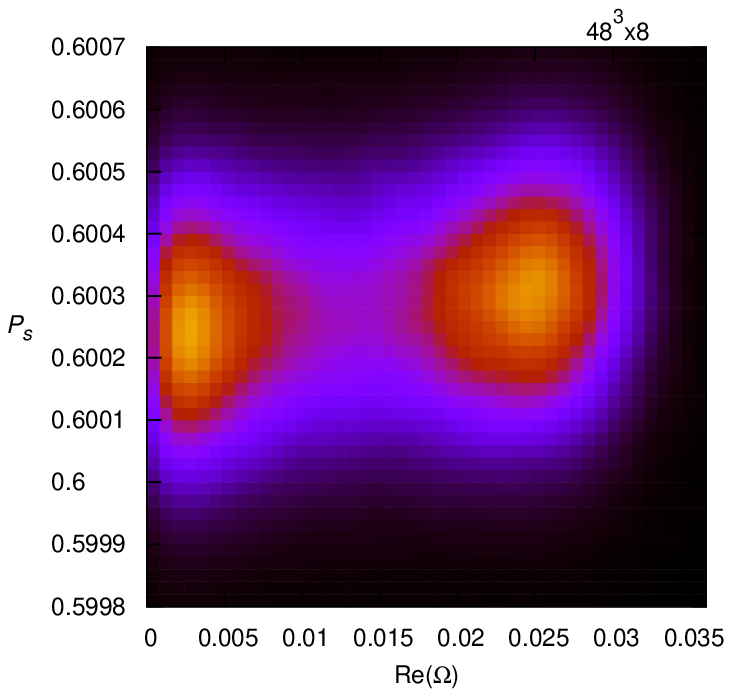}
\includegraphics[width=80mm]{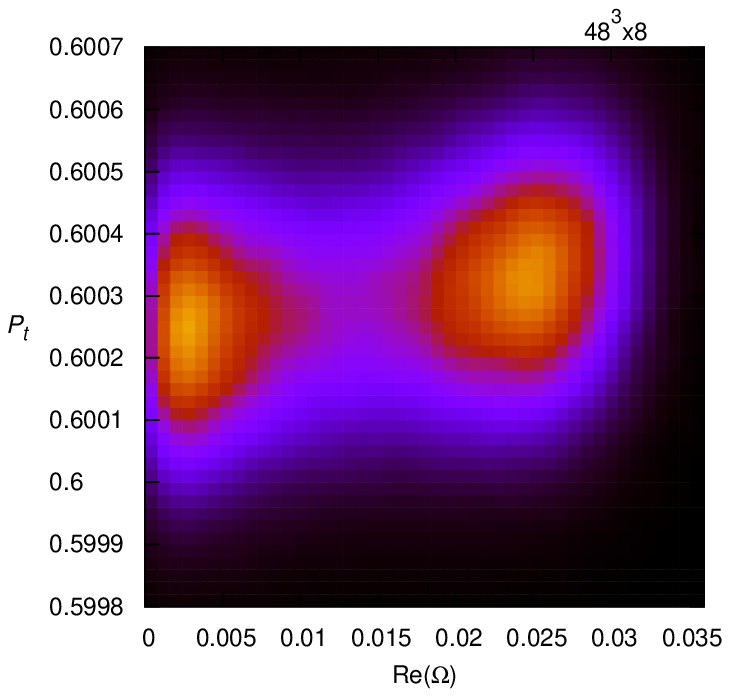} \\
\includegraphics[width=80mm]{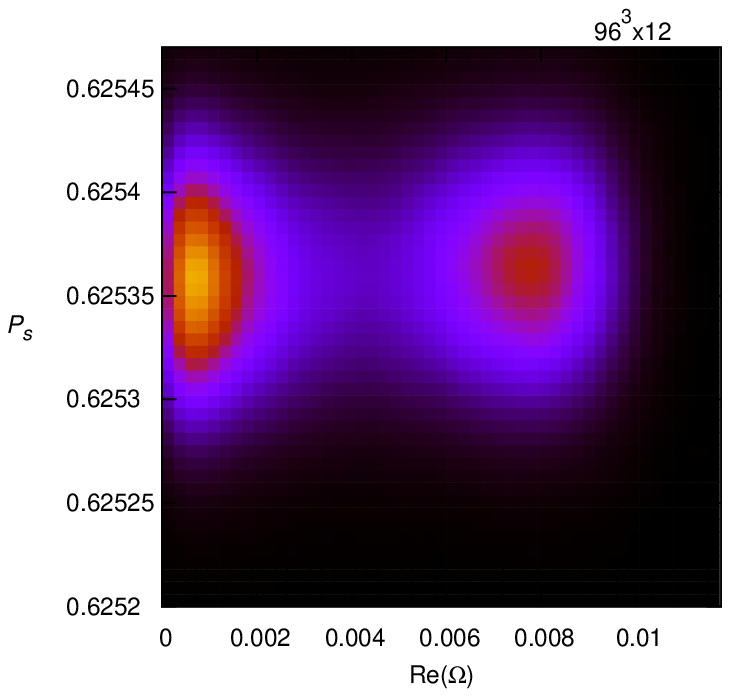}
\includegraphics[width=80mm]{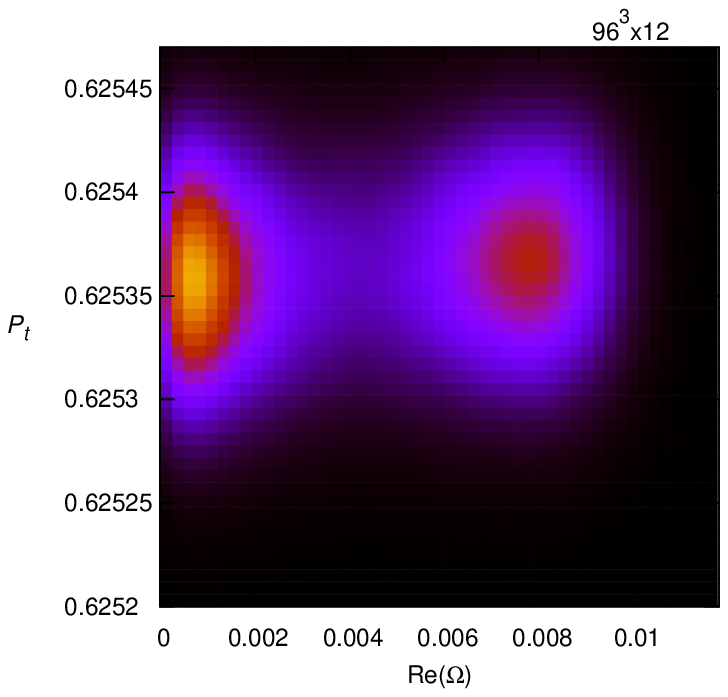}
\caption{Histograms as functions of $(P_s, {\rm Re} \Omega)$ (left) and $(P_t, {\rm Re} \Omega)$ (right)  at the transition point,
obtained on the $48^3 \times 8$ (top) and $96^3 \times 12$ (bottom) lattices.
Brighter color means larger probability.
}
\label{fig:histo}
\end{figure}

\begin{figure}[tb]
\centering
\includegraphics[width=80mm]{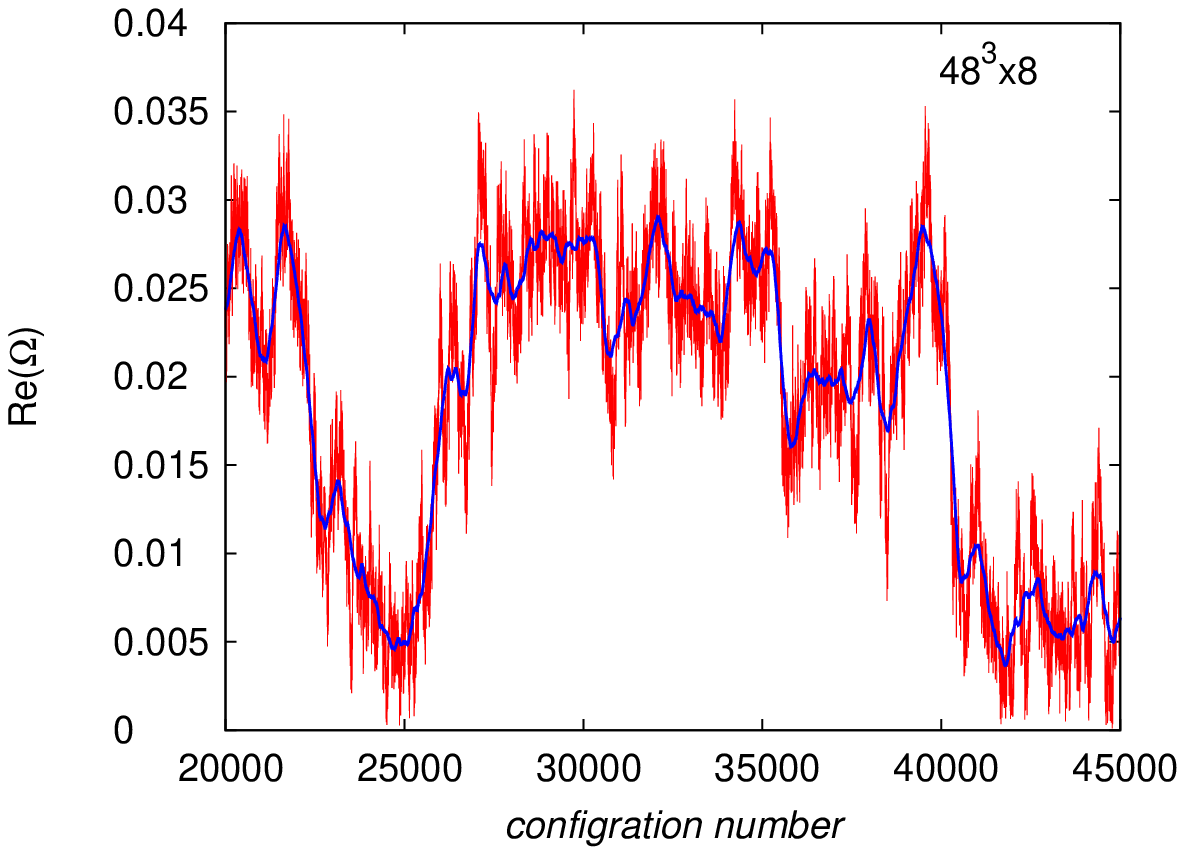}
\includegraphics[width=80mm]{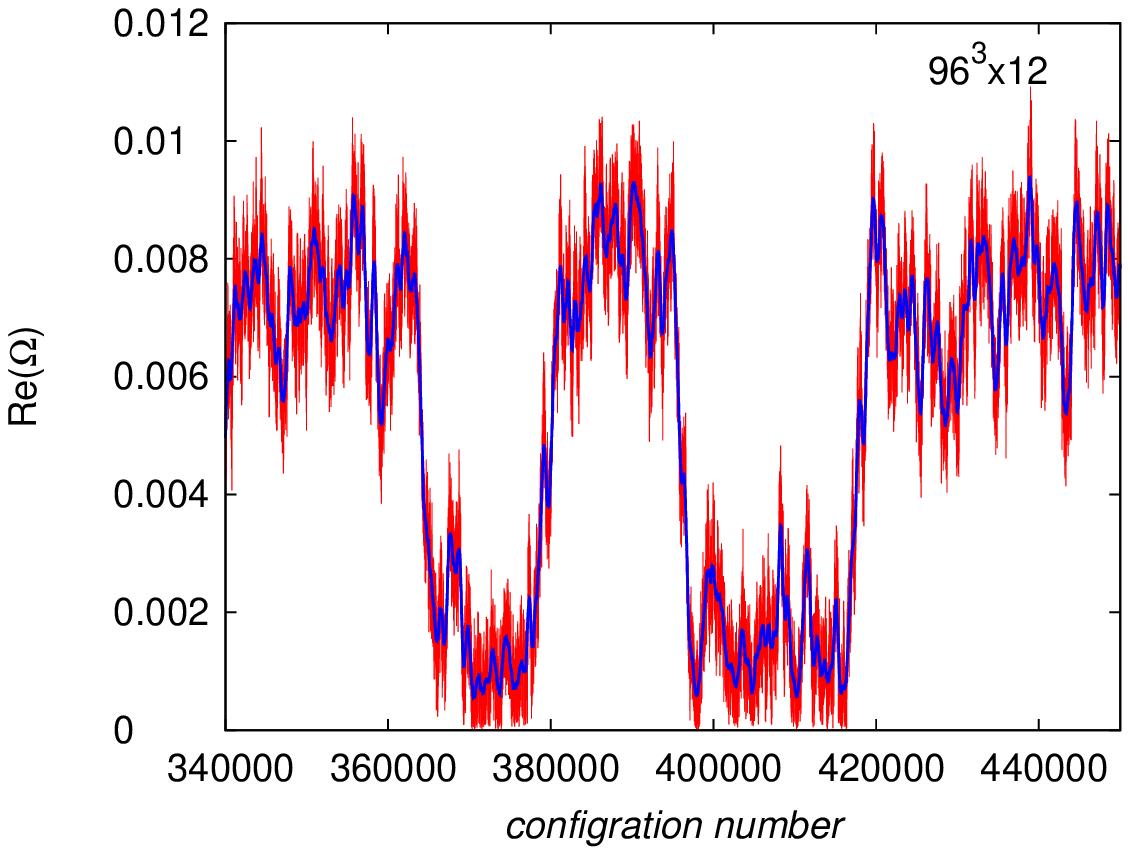}
\caption{A part of the time history of Polyakov loop ${\rm Re}\Omega$ on the $48^3 \times 8$ lattice at $\beta=6.020$ (left) and $96^3\times12$ at $\beta=6.335$ (right).
The solid blue curves show the smeared Polyakov loop over $\pm250$ configurations around the current configuration number.
}
\label{fig:timehis}
\end{figure}

\begin{figure}[tb]
\centering
\includegraphics[width=80mm]{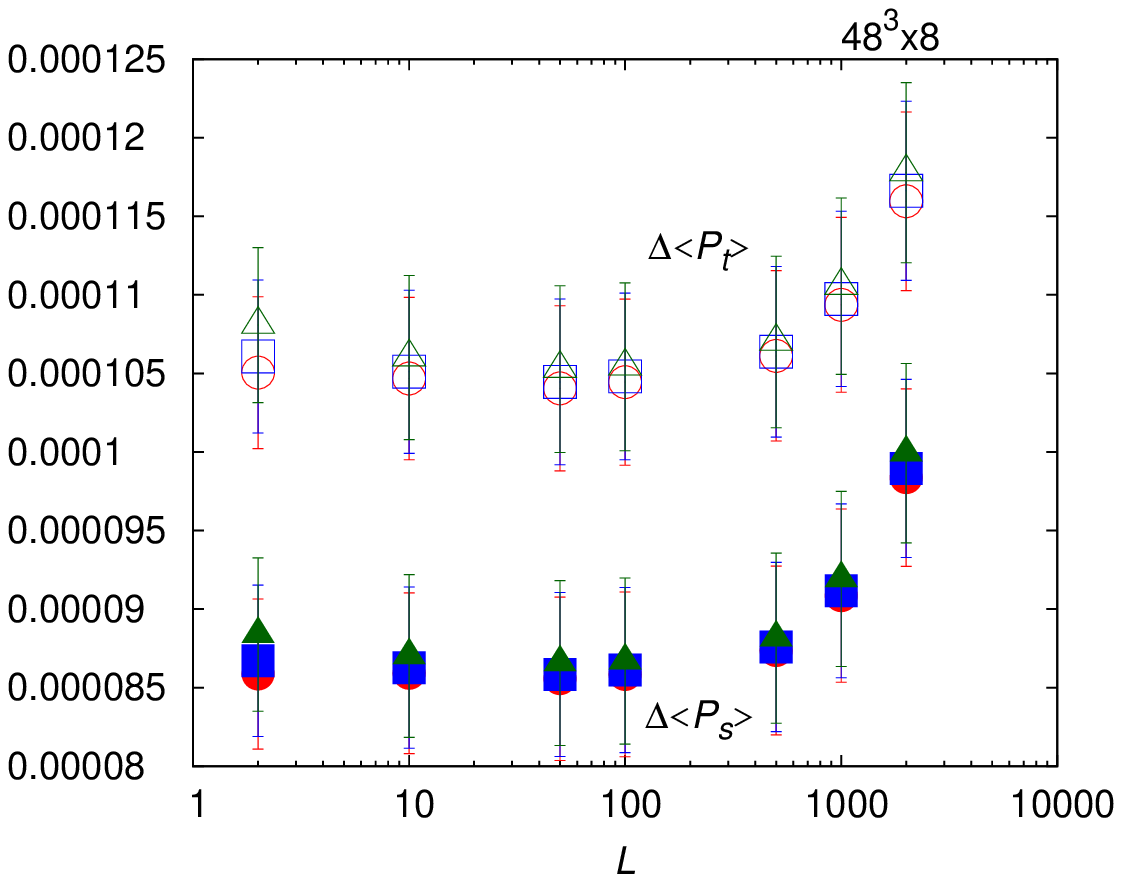}
\includegraphics[width=80mm]{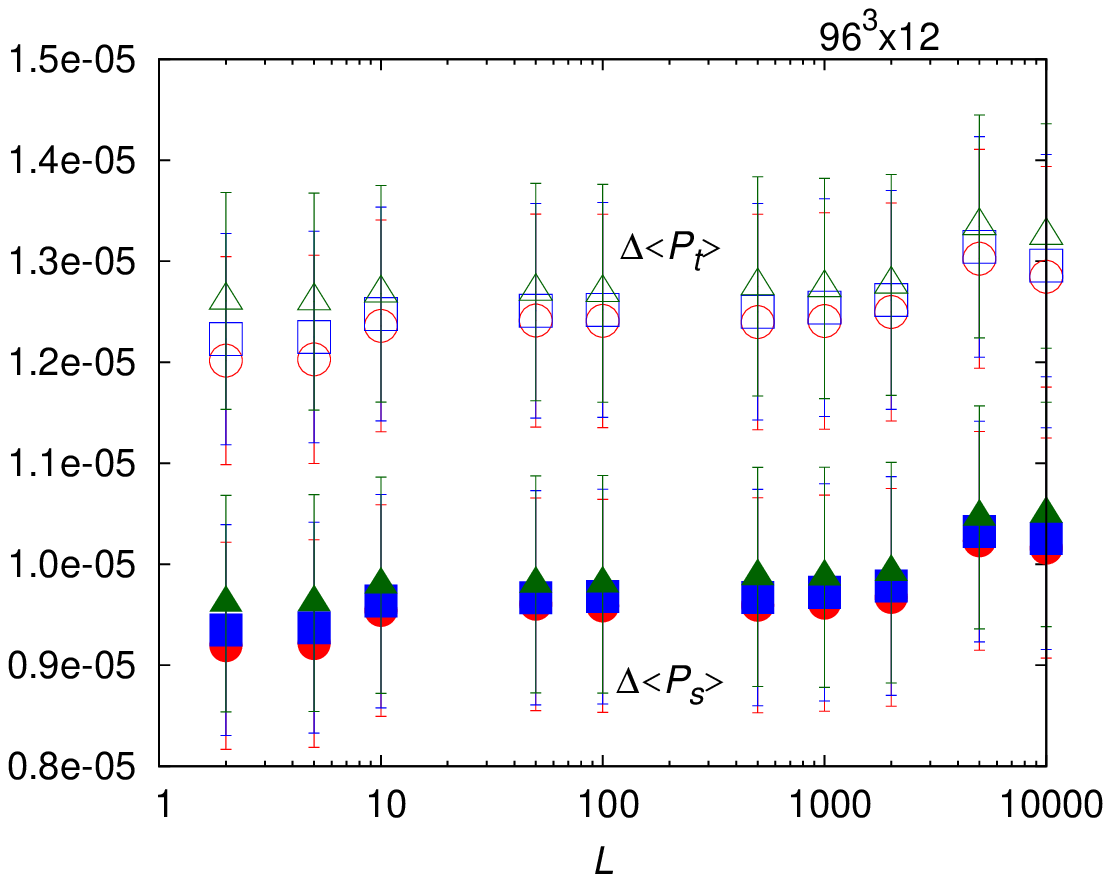}
\caption{Systematic uncertainties in the gap of space-like and time-like plaquette 
$\Delta \langle P_s \rangle$ and $\Delta \langle P_t \rangle$ depending on the way of the phase separation, 
obtained on $48^3\times 8$ (left) and $96^3\times12$ (right) lattices.
Circular symbols are obtained by separating the configurations at minimum of ${\rm Re}\Omega$ histogram between the two peaks.
Square symbols are the results of our choice of the thresholds LB(hot) and UB(cold) given in Table ~\ref{tab:polyakov}, and triangular symbols are the results adopting three times wider gaps between LB(hot) and UB(cold).
}
\label{fig:comp}
\end{figure}

\begin{table}[tb]
\caption{Expectation values of Polyakov loop at the transition point in the hot and cold phases.  
Also given are the thresholds for phase separation: lower bound for the hot phase (LB(hot)) and upper bound for the cold phase (UB(cold)).
}
\label{tab:polyakov}
\begin{tabular}{lllcc}
\hline
\hline
lattice       &$\langle\Omega\rangle_{\rm hot}$&$\langle\Omega\rangle_{\rm cold}$&LB(hot)&UB(cold)\\\hline
$48^3\times6$ &0.04485(54) &0.00550(16) &0.0249&0.0231\\
$64^3\times6$ &0.04568(37) &0.00697(25) &0.0241&0.0239\\
$48^3\times8$ &0.02312(18) &0.00640(15) &0.0138&0.0136\\
$64^3\times8$ &0.02102(12) &0.004769(56)&0.0126&0.0124\\
$64^3\times12$&0.007412(90)&0.002075(53)&0.0039&0.0037\\
$96^3\times12$&0.007256(79)&0.001482(48)&0.0039&0.0037\\
\hline
\hline
\end{tabular}
\end{table}

\begin{table}[tb]
\caption{Expectation values of space-like and time-like plaquettes at the transition point 
in the hot and cold phases,
and their gaps.
}
\label{tab:plaquette}
\begin{tabular}{cllllll}
\hline
\hline
lattice
&$\left\langle P_s\right\rangle_{\rm cold}$ & $\left\langle P_t\right\rangle_{\rm cold}$
&$\left\langle P_s\right\rangle_{\rm hot}$ & $\left\langle P_t\right\rangle_{\rm hot}$
&$\Delta \left\langle P_s\right\rangle$
&$\Delta \left\langle P_t\right\rangle$ \\
\hline
$48^3\times6$ &0.58111347(96)&0.5811273(10) &0.5814088(76) &0.5814875(91) &0.0002954(76)&0.0003602(92)\\
$64^3\times6$ &0.58119760(23)&0.58121330(23)&0.58151940(47)&0.58160140(50)&0.0003215(71)&0.0003876(78)\\
$48^3\times8$ &0.6002544(39) &0.6002574(38) &0.6003397(26) &0.6003617(31) &0.0000912(55)&0.0001097(56)\\
$64^3\times8$ &0.6002977(15) &0.6003163(16) &0.60023734(89)&0.60024163(90)&0.0000612(17)&0.0000755(18)\\
$64^3\times12$&0.6253174(36) &0.6253176(36) &0.6253256(22) &0.6253293(22) &0.0000103(24)&0.0000137(25)\\
$96^3\times12$&0.6253577(12) &0.6253584(12) &0.6253674(12) &0.6253710(12) &0.0000096(11)&0.0000124(11)\\
\hline
\hline
\end{tabular}
\end{table}

To evaluate the latent heat and the pressure gap, we need to separate the configurations at the first order transition point into 
the hot and cold phases.
From Eqs.~(\ref{eq:denrg}) and (\ref{eq:dprs}), $\Delta \epsilon /T^4$ 
and $\Delta p /T^4$ are proportional to $N_t^4$ and thus the gaps in the plaquettes will decrease as $1/N_t^4$ near the continuum limit. 
This indicates that a high precision measurement is required at large $N_t$.

In Fig.~\ref{fig:histo} we show some contour plots of the histograms as functions of $(P_s, {\rm Re} \Omega)$ and $(P_t, {\rm Re} \Omega)$, 
obtained on $48^3 \times 8$ and $96^3 \times 12$ lattices.
Using the multi-point reweighting method, $\beta$ is adjusted to the transition point. 
The two peaks correspond to the hot and cold phases. 
The peaks are well separated in the ${\rm Re}\Omega$ direction, while they are overlapping in the plaquette directions.

To separate the two phases, we introduce cuts in the time history of the Polyakov loop \cite{FOU,QCDPAX}.
The red lines in Fig.~\ref{fig:timehis} show a part of the time history of 
the Polyakov loop, enlarged around a flip-flop, obtained on the $48^3 \times 8$ and $96^3 \times 12$ lattices.
The former lattice is an example on which the phase separation is relatively difficult.
To remove jagged fluctuations, we average ${\rm Re} \Omega$ over $\pm 250$ configurations around the current configuration number.  
The total number of configurations to be averaged, which we call the smearing width, is 501 in this case. 
The results of the time-smeared Polyakov loop are shown by the solid blue curves in Fig.~\ref{fig:timehis} .
We then identify the hot/cold phase when the time-smeared Polyakov loop is larger/smaller than a lower/upper bound value. 
The configurations with the time-smeared Polyakov loop between the thresholds are discarded as the mixed phase. 
The values of the thresholds we adopt are given in Table~\ref{tab:polyakov}.
We show in the following that these choices give stable gaps on our lattices.

After separating out the two phases at each simulation point, we combine the configurations by the multi-point reweighting method to compute expectation values in each phase just at the transition point.
In Table~\ref{tab:polyakov}, the results of Polyakov loop expectation values in each phase at $\beta_c(N_t,V)$ are given.
The results of $\langle P_{s/t} \rangle_{\rm hot/cold}$
as well as the plaquette gaps, 
$\Delta \langle P_{s/t} \rangle \equiv \langle P_{s/t} \rangle_{\rm hot} - \langle P_{s/t} \rangle_{\rm cold}$, 
are summarized in Table \ref{tab:plaquette}.

To estimate systematic uncertainty due to the phase separation procedure, we repeat the study by varying the smearing width and the values for the thresholds. 
The smearing width $L$ may be longer than the persistent time of the mixed phase, but should be much smaller than the persistent time of hot and cold phases.
In Fig.~\ref{fig:comp}, we plot the results of 
$\Delta \langle P_s \rangle$ and $\Delta \langle P_t \rangle$ 
computed on the $48^3 \times 8$ and $96^3\times12$ lattices near the transition point 
as functions of $L$.
Filled symbols and open symbols are $\Delta \langle P_s \rangle$ and $\Delta \langle P_t \rangle$, respectively.
Circular symbols are obtained by separating the configurations at the minimum of ${\rm Re}\Omega$ histogram between the two peaks.
Square symbols are the results with the thresholds given in Table~\ref{tab:polyakov}.
Triangular symbols the results with three times wider gaps between the thresholds than our choices given in Table~\ref{tab:polyakov}.
We see that, by removing the mixed phase configurations between the thresholds, the gaps becomes slightly larger, but the shift is much smaller than the statistical errors. 
This means that the dependence on the values of the thresholds is negligible on our lattices.
We also see that the results are quite stable for $L \simle 501$. 
From the stability of the results, we adopt $L=501$ in our study. 
\footnote{
In Refs.~\cite{QCDPAX}, a more elaborated method was adopted to remove the contributions of the mixed phase: 
Sufficient number of configurations around the flip-flop points are removed until the results for $\langle P_{s/t} \rangle_{\rm hot/cold}$ become stable.
However, because the spatial lattice sizes are much larger in our study, we expect that the contribution of the mixed phase is much smaller on our lattices.
Actually, dependences on the choice of the threshold and the smearing width are very small, as discussed in the text.
We thus adopt the simpler method in our multi-point analyses.
}

\subsection{Latent heat and pressure gap}
\label{sec:latent}

\begin{figure}[tb]
\centering
\includegraphics[width=90mm]{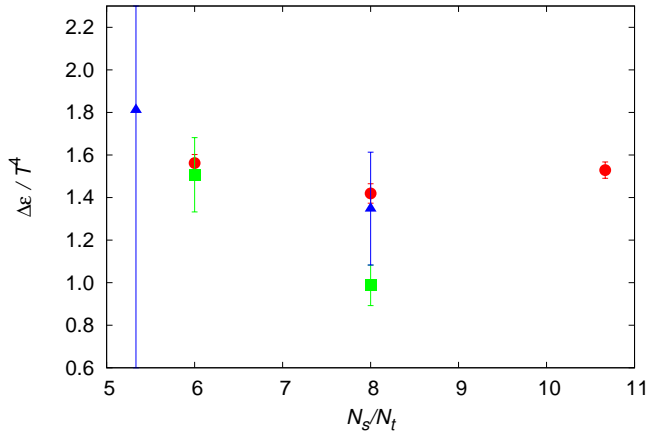}
\caption{Latent heat $\Delta \epsilon /T^4$ for $N_t=6$ (circle), $8$ (square) and $12$ (triangle) as a function of the aspect ratio $N_s/N_t$.
}
\label{fig:vdep}
\end{figure}

\begin{figure}[tb]
\centering
\includegraphics[width=90mm]{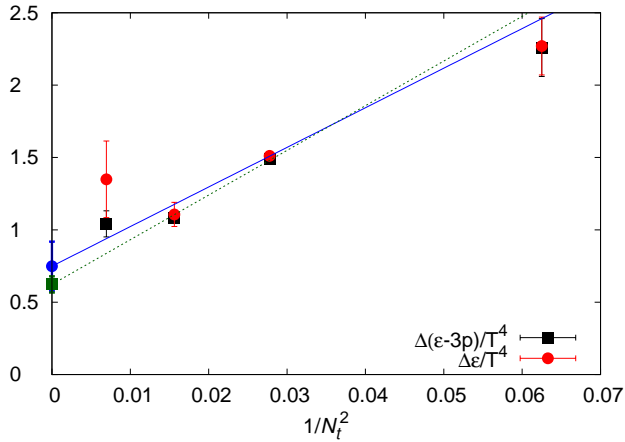}
\caption{Continuum extrapolation of the latent heat $\Delta \epsilon/T^4$ (circle) and  $\Delta (\epsilon -3 p)/T^4$ (square) using data at $N_t=6$, 8 and 12.
The rightmost data at $N_t=4$ are obtained in Ref.~\cite{ejiri98}, with error bars including systematic errors due to the choice of the beta function. See text for details.
}
\label{fig:latent}
\end{figure}

\begin{table}
\caption{Gaps of thermodynamic quantities at the deconfining transition point of the SU(3) gauge theory using non-perturbative anisotropy coefficients. }
\label{tab:dep}
\begin{tabular}{llllll}
\hline
\hline
lattice&$\beta_c(N_t,V)$&$\Delta \epsilon /T^4$&$\Delta p / T^4$&$\Delta (\epsilon + p)/T^4$
&$\Delta (\epsilon -3 p)/T^4$\\
\hline
$36^2\times48\times6$&5.89379(34) &1.56(4)  &-0.003(17)&1.56(5)  &1.57(4)\\
$48^3\times6$        &5.89383(24) &1.42(5)  &0.007(11) &1.43(5)  &1.40(4)\\
$64^3\times6$        &5.894512(40)&1.53(4)  &0.006(7)  &1.53(4)  &1.51(3)\\
$48^3\times8$        &6.06160(18) &1.51(17) &0.009(43) &1.52(21) &1.48(8)\\
$64^3\times8$        &6.06247(14) &0.99(10) &-0.02(3)  &0.97(12) &1.04(3)\\
$64^3\times12$       &6.3349(11)  &1.81(509)&0.24(172) &2.05(681)&1.09(21)\\
$96^3\times12$       &6.33532(11) &1.35(27) &0.11(8)   &1.45(34) &1.03(10)\\
\hline
\hline
\end{tabular}
\end{table}

\begin{table}
\caption{Latent heat in the thermodynamic limit at each $N_t$.}
\label{tab:deNt}
\begin{tabular}{cll}
\hline
\hline
$N_t$&$\Delta\epsilon/T^4$&$\Delta(\epsilon-3p)/T^4$\\
\hline
6 &1.511(24) &1.488(21)\\
8 &1.106(84) &1.079(25)\\
12&1.349(265)&1.041(90)\\
\hline
\hline
\end{tabular}
\end{table}

\begin{table}
\caption{Latent heat in the continuum limit.}
\label{tab:deCont}
\begin{tabular}{ccccc}
\hline
\hline
fit range &\multicolumn{2}{c}{$N_t=6$--12} & \multicolumn{2}{c}{$N_t=4$--12} \\
 &$\Delta\epsilon/T^4$&$\Delta(\epsilon-3p)/T^4$  &$\Delta\epsilon/T^4$&$\Delta(\epsilon-3p)/T^4$ \\
\hline
 &0.75(17) & 0.623(56) &0.83(12)&0.652(51) \\
$\chi^2/$dof  & 3.12 & 6.38 & 1.75 & 4.05 \\
\hline
\hline
\end{tabular}
\end{table}

Using the non-perturbative anisotropy coefficients (Table~\ref{tab:cst}) and the plaquette gaps (Table~\ref{tab:plaquette}), 
we compute the latent heat $\Delta \epsilon$ and the pressure gap $\Delta p$ using Eqs.~(\ref{eq:denrg}) and (\ref{eq:dprs}).
For the lattice $36^2\times48\times6$, we adopt the results of plaquette gaps by Ref.~\cite{QCDPAX} and the anisotropy coefficients given in Table~\ref{tab:cst}.
The results of latent heat and the pressure gap using our non-perturbative 
anisotropy coefficients are summarized in Table~\ref{tab:dep}.

It is known that the perturbative anisotropy coefficients lead to a 
difficulty of non-vanishing pressure gaps, $\Delta p / T^4 = -0.32(3)$ and $-0.14(2)$ at
$N_t=4$ and 6 \cite{QCDPAX}.
From Table~\ref{tab:dep}, we find that the problem is completely resolved with non-perturbative anisotropy coefficients. 
We also confirm that the condition for vanishing pressure gap, 
Eq.~(\ref{eq:dp0cond}), is well satisfied on each of our lattices, as shown in Table~\ref{tab:rt}.

We now study the spatial volume dependence of the results.
Because the correlation length remains finite at first order phase transition point, when the spatial lattice size is sufficiently larger than the correlation length, the gaps will saturate. 
Thus we expect that $\Delta \epsilon$ at the deconfining transition point is independent of the spatial volume on sufficiently large lattices.
In Fig.~\ref{fig:vdep}, we plot $\Delta \epsilon/T^4$ at $N_t=6$ (circle), $8$ (square) and $12$ (triangle) as a function of the aspect ratio.
From the $N_t=6$ results, we find that the latent heat is well stable at $N_s/N_t \simge 6$.
The results at $N_t=8$ and 12 are also consistent with constant, although the errors are larger.
We thus perform a constant fit of the data shown in Fig.~\ref{fig:vdep} at each $N_t$. 
The results of the constant fit at each $N_t$ are given in Table~\ref{tab:deNt}, together with the results of similar analysis for $\Delta(\epsilon-3p)$.
Because the anisotropy coefficients are not needed for $\Delta (\epsilon -3p)/T^4$, the statistical errors are smaller than those of $\Delta \epsilon/T^4$.

Finally, we extrapolate the results to the continuum limit.
Because the leading lattice artifact in the action is $O(a^2)$ and also the equation of state in the high temperature limit is a function of $N_t^2$, we carry out linear extrapolations in $1/N_t^2$. 
Using the data at $N_t =6$, 8 and 12,  we obtain the solid and dashed lines in Fig.~\ref{fig:latent} 
which give 
\begin{eqnarray}
\Delta \epsilon/T^4 = 0.75 \pm 0.17
,\hspace{5mm}
\Delta (\epsilon -3 p)/T^4 = 0.623 \pm 0.056
\end{eqnarray}
in the continuum limit, Their $\chi^2/$dof are given in Table~\ref{tab:deCont}.

In Fig.~\ref{fig:latent}, we also show the results at $N_t=4$ obtained in Ref.~\cite{ejiri98} using non-perturbative anisotropy coefficients \cite{ejiri98} and beta function defined by $\beta_c$ \cite{boyd}. 
As discussed at the end of Sec.~\ref{sec:aniso}, uncertainty due to the definition of the beta function is large at $N_t=4$.
We thus include the systematic error from the definition of the beta function (estimated as half of the maximum difference among three beta functions) to the error bars for these data. 
Because the data at $N_t=4$ turned out to be not far from the fitting lines in Fig.~\ref{fig:latent}, we also tried fits including the data at $N_t=4$.
The results are given in the last two columns in Table~\ref{tab:deCont}.
We see that the results are stable under the change of the fitting range, though the errors are not quite small yet.

\section{Conclusion and outlook}
\label{sec:conclusion}

Performing a series of finite temperature simulations of the SU(3) gauge theory on isotropic $N_t=6$, 8 and 12 lattices with the aspect ratio $N_s/N_t = 5.3$--10.7, 
we computed non-perturbative values of the anisotropy coefficients at the first order deconfining phase transition point  
by measuring the transition line in 
the $(\beta_s, \beta_t)$ plane using the multi-point reweighting method.
We found that the non-perturbative anisotropy coefficients approach their perturbative values as 
increasing $N_t$.

We then computed the gaps of several observables
between the high and low temperature phases at the first order transition point, 
by separating out the configurations in each phases. 
From the results of the non-perturbative anisotropy coefficients and the plaquette gaps, 
we calculated the latent heat and the pressure gap at the transition point. 
We confirmed that the non-perturbative anisotropy coefficients lead to vanishing pressure gaps on our finite lattices.
Studying the spatial volume dependence and carrying out the continuum extrapolation, 
the latent heat was found to be 
$
\Delta \epsilon/T^4 = 0.75 \pm 0.17
$
and
$
\Delta (\epsilon -3 p)/T^4 = 0.623 \pm 0.056
$
in the continuum limit.

Our direct calculation of the anisotropy coefficients suggests that the perturbative values  of the anisotropy coefficients may be recovered at $N_t \simge 12$, though the errors are not small yet.
If this is so, the derivative method with perturbative anisotropy coefficients is applicable around the transition point when $N_t \simge 12$.
This may help calculation of the equation of state in full QCD, 
where a precise evaluation of $(\epsilon -3p)/T^4$ needed in the integral method is quite costly at low temperatures (large $N_t$). 
The derivative method is an attractive choice also in the fixed scale approach \cite{umeda12,umeda15}.

\section*{Acknowledgments}
We would like to thank other members of the WHOT-QCD Collaboration for discussions.
This work is in part supported by JSPS KAKENHI Grant
No.\ 25800148, No.\ 26287040, No.\ 26400244, No.\ 26400251, and No.\ 15K05041, and by the 
Large Scale Simulation Program of High Energy Accelerator
Research Organization (KEK) No.\ 14/15-23, 15/16-T06, 15/16-T-07, and 15/16-25.



\begin{thebibliography}{99}

\bibitem{YHM}
See, e.g., K. Yagi, T. Hatsuda, and Y. Miake, 
Quark-Gluon Plasma (Cambridge University Press, Cambridge, 2005).

\bibitem{Appelquist:1995en}
T.~Appelquist, M.~Schwetz and S.~B.~Selipsky,
A strongly first order electroweak phase transition from strong 
symmetry-breaking interactions, 
Phys.~Rev.~D {\bf 52}, 4741 (1995).

\bibitem{Kikukawa:2007zk} 
Y.~Kikukawa, M.~Kohda, and J.~Yasuda,
First-order restoration of ${\rm SU}(N_f) \times {\rm SU}(N_f)$ chiral 
symmetry with large $N_f$ and Electroweak phase transition, 
Phys.~Rev.~D {\bf 77}, 015014 (2008).

\bibitem{Ejiri:2012rr}
S.~Ejiri and N.~Yamada,
End Point of a First-Order Phase Transition in Many-Flavor Lattice
QCD at Finite Temperature and Density, 
Phys.~Rev.~Lett.~{\bf 110}, 172001 (2013).

\bibitem{integral} 
J.~Engels, J.~Fingberg, F.~Karsch, D.~Miller, M.~Weber,
Nonperturbative thermodynamics of SU(N) gauge theories,
Phys.~Lett.~B {\bf 252}, 625 (1990).

\bibitem{karsch} 
F.~Karsch, 
SU(N) Gauge Theory Couplings on Asymmetric Lattices, 
Nucl.~Phys.~B {\bf 205}, 285 (1982).

\bibitem{ejiri98}
S.~Ejiri, Y.~Iwasaki and K.~Kanaya, 
Nonperturbative determination of anisotropy coefficients 
in lattice gauge theories, 
Phys.~Rev.~D {\bf 58}, 094505 (1998).

\bibitem{satz} 
J.~Engels, F.~Karsch, H.~Satz and I.~Montvay, 
Gauge Field Thermodynamics for the SU(2) Yang-Mills System,
Nucl.~Phys.~B {\bf 205}, 545 (1982).

\bibitem{burgers} 
G.~Burgers, F.~Karsch, A.~Nakamura and I.O.~Stamatescu, 
QCD on anisotropic lattices, 
Nucl.~Phys.~B {\bf 304}, 587 (1988).

\bibitem{fujisaki}
QCD-TARO Collaboration: M.~Fujisaki {\it et al.}, 
Finite temperature gauge theory on anisotropic lattices, 
Nucl.~Phys.~B(Proc.~Suppl.) {\bf 53}, 426 (1997).

\bibitem{scheideler} 
J.~Engels, F.~Karsch and T.~Scheideler,
Direct determination of the gauge coupling derivatives for 
the energy density in lattice QCD,
Nucl.~Phys.~B(Proc.~Suppl.) {\bf 63}, 427 (1998). 

\bibitem{engels00} 
J.~Engels, F.~Karsch and T.~Scheideler, 
Determination of anisotropy coefficients for SU(3) gauge actions 
from the integral and matching methods, 
Nucl.~Phys.~B {\bf 564}, 303 (2000).

\bibitem{klassen}
T.R.~Klassen, 
The Anisotropic Wilson gauge action, 
Nucl.~Phys.~B {\bf 533}, 557 (1998). 

\bibitem{boyd} 
G.~Boyd, J.~Engels, F.~Karsch, E.~Laermann, C.~Legeland, 
M.~Lutgemeier, B.~Petersson, 
Thermodynamics of SU(3) lattice gauge theory, 
Nucl.~Phys.~B {\bf 469}, 419 (1996).

\bibitem{edwards}
R.G.~Edwards, U.M.~Heller and T.R.~Klassen, 
Accurate scale determinations for the Wilson gauge action, 
Nucl.~Phys.~B {\bf 517}, 377 (1998).

\bibitem{taro} 
K.~Akemi {\it et al.} (QCD-TARO Collaboration), 
Scaling study of pure gauge lattice QCD by Monte Carlo 
renormalization group method, 
Phys.~Rev.~Lett.~{\bf 71}, 3063 (1993).

\bibitem{Guagnelli:1998ud}
M.~Guagnelli, R.~Sommer and H.~Wittig [ALPHA Collaboration],
Precision computation of a low-energy reference scale in quenched lattice QCD,
Nucl.~Phys.~B {\bf 535}, 389 (1998).

\bibitem{Necco:2001xg}
S.~Necco and R.~Sommer,
The $N_f = 0$ heavy quark potential from short to intermediate distances,
Nucl.~Phys.~B {\bf 622}, 328 (2002).

\bibitem{Asakawa:2015vta}
M.~Asakawa, T.~Hatsuda, T.~Iritani, E.~Itou, M.~Kitazawa and H.~Suzuki,
Determination of Reference Scales for Wilson Gauge Action from 
Yang--Mills Gradient Flow,
arXiv:1503.06516 [hep-lat].

\bibitem{francis15} 
A.~Francis, O.~Kaczmarek, M.~Laine, T.~Neuhaus, H.~Ohno, 
Critical point and scale setting in SU(3) plasma: An update, 
Phys.~Rev.~D {\bf 91}, 096002 (2015).

\bibitem{ejiri03}
S.~Ejiri, 
Remarks on the multiparameter reweighting method for the study of 
lattice QCD at nonzero temperature and density, 
Phys.~Rev.~D {\bf 69}, 094506 (2004).

\bibitem{MDS67}
I.R.~McDonald and K.~Singer,
Calculation of thermodynamic properties of liquid argon from Lennard-Jones parameters by a Monte Carlo method,
Discuss.\ Faraday Soc.~{\bf 43}, 40 (1967).

\bibitem{FS89} 
A.M.~Ferrenberg and R.H.~Swendsen, 
New Monte Carlo technique for studying phase transitions
Phys.~Rev.~Lett.~{\bf 61}, 2635 (1988);
Optimized Monte Carlo analysis, 
Phys.~Rev.~Lett.~{\bf 63}, 1195 (1989).

\bibitem{iwami15}
R.~Iwami, S.~Ejiri, K.~Kanaya, Y.~Nakagawa, D.~Yamamoto, T.~Umeda 
(WHOT-QCD Collaboration), 
Multipoint reweighting method and its applications to lattice QCD, 
Phys.~Rev.~D {\bf 92}, 094507 (2015).

\bibitem{QCDPAX}
Y.~Iwasaki, K.~Kanaya, T.~Yoshi\'e, T.~Hoshino, T.~Shirakawa, 
Y.~Oyanagi, S.~Ichii, and T.~Kawai,
Deconfining transition of SU(3) gauge theory on $N_t=4$ and 6 lattices,
Phys.~Rev.~Lett.~{\bf 67}, 3343 (1991);
Finite temperature phase transition of SU(3) gauge theory 
on $N_t =4$ and 6 lattices,
Phys.~Rev.~D {\bf 46}, 4657 (1992).

\bibitem{FOU}
M.~Fukugita, M.~Okawa and A.~Ukawa,
Order of the deconfining phase transition in SU(3) lattice gauge theory, 
Phys.~Rev.~Lett.~{\bf 63}, 1768 (1989);
Finite Size Scaling Study of the Deconfining Phase Transition 
in Pure SU(3) Lattice Gauge Theory,
Nucl.~Phys.~B {\bf 337}, 181 (1990).

\bibitem{umeda12}
T.~Umeda, S.~Aoki, S.~Ejiri, T.~Hatsuda, K.~Kanaya, Y.~Maezawa, H.~Ohno 
(WHOT-QCD Collaboration), 
Equation of state in 2+1 flavor QCD with improved Wilson quarks 
by the fixed scale approach, 
Phys.~Rev.~D {\bf 85}, 094508 (2012).

\bibitem{umeda15}
T.~Umeda, S.~Ejiri, R.~Iwami, K.~Kanaya (WHOT-QCD Collaboration), 
Towards the QCD equation of state at the physical point using Wilson fermion, 
Proc.~of~Sci.~(LATTICE 2015) 209 (2015), arXiv:1511.04649.


\end{thebibliography}
\end{document}